\documentclass[12pt]{article}
\usepackage{amsmath}
\usepackage{amssymb}
\usepackage{graphicx}
\usepackage{ifthen}
\usepackage{verbatim}

\newlength{\captionwidth}
\newsavebox{\tempbox}
\newcommand{\mycaption}[2]{%
\par\vspace{10pt}\sbox{\tempbox}{Figure #1: #2}%
\ifthenelse{\lengthtest{\wd\tempbox>\captionwidth}}%
{\sbox{\tempbox}{Figure.#1:\ }%
\addtolength{\captionwidth}{-\wd\tempbox}%
\mbox{Figure #1:\ }\parbox[t]{\captionwidth}{\small\textit{#2}}}%
{Figure #1: {\small\textit{#2}}}}%

\makeatletter
  
  \@addtoreset{equation}{section}
\makeatother

\allowdisplaybreaks
\begin{document}
\thispagestyle{empty}
\begin{flushright}
\texttt{hep-th/0412329}\\
OU-HET 513 \\
December 2004
\end{flushright}
\bigskip
\bigskip
\begin{center}
{\Large \textbf{Free Fermion and Seiberg-Witten Differential}}
\end{center}
\begin{center}
{\Large \textbf{in Random Plane Partitions}}
\end{center}
\bigskip
\bigskip
\renewcommand{\thefootnote}{\fnsymbol{footnote}}
\begin{center}
Takashi Maeda\footnote{E-mail: 
\texttt{maeda@het.phys.sci.osaka-u.ac.jp}}$^1$, 
Toshio Nakatsu\footnote{E-mail: 
\texttt{nakatsu@het.phys.sci.osaka-u.ac.jp}}$^1$, 
Kanehisa Takasaki\footnote{E-mail: 
\texttt{takasaki@math.h.kyoto-u.ac.jp}}$^2$
and 
Takeshi Tamakoshi\footnote{E-mail: 
\texttt{tamakoshi@het.phys.sci.osaka-u.ac.jp}}$^1$\\
\bigskip
{\small 
\textit{$^1$Department of Physics, Graduate School of Science, 
Osaka University,\\
Toyonaka, Osaka 560-0043, Japan\\
$^2$Graduate School of Human and Environmental Studies, 
Kyoto University, \\
Yoshida, Sakyo, Kyoto 606-8501, Japan
}}
\end{center}
\bigskip
\bigskip
\renewcommand{\thefootnote}{\arabic{footnote}}
\begin{abstract}
A model of random plane partitions which describes  
five-dimensional $\mathcal{N}=1$ supersymmetric 
$SU(N)$ Yang-Mills is studied. 
We compute the wave functions of fermions 
in this statistical model and investigate 
their thermodynamic limits or the semi-classical behaviors. 
These become of the WKB type at the thermodynamic limit. 
When the fermions are located at the main diagonal 
of the plane partition, their semi-classical wave functions 
are obtained in a universal form. 
We further show that 
by taking the four-dimensional limit 
the semi-classical wave functions turn to live 
on the Seiberg-Witten curve and that the classical action 
becomes precisely the integral of the Seiberg-Witten differential. 
When the fermions are located away from the main diagonal, 
the semi-classical wave functions depend on 
another continuous parameter.  
It is argued that they are related with the wave functions 
at the main diagonal by the renormalization group 
flow of the underlying gauge theory. 
\end{abstract}

\setcounter{footnote}{0}
\newpage
%
%
\section{Introduction}

Recently it becomes possible 
\cite{Nekrasov,Nekrasov-Okounkov} to compute 
the exact partition functions of four-dimensional 
$\mathcal{N}=2$ supersymmetric gauge theories. 
The celebrated Seiberg-Witten solutions 
\cite{Seiberg-Witten} of the gauge theories emerge 
\cite{Nekrasov-Okounkov,Nakajima-Yoshioka} through the statistical 
models of random partitions. 
In particular, Nekrasov and Okounkov show 
\cite{Nekrasov-Okounkov} that the Seiberg-Witten 
geometry is realized as the thermodynamic limit  
or the semi-classical approximation of the statistical 
models.

In \cite{MNTT} the authors consider 
the statistical model of random plane partitions 
relevant to describe five-dimensional 
$\mathcal{N}=1$ supersymmetric $SU(N)$ Yang-Mills theory 
on $\mathbb{R}^4 \times S^1$ 
(more precisely plus the Chern-Simons term). 
This model has a smooth four-dimensional limit and 
gives rise to the exact partition function for 
four-dimensional $\mathcal{N}=2$ supersymmetric Yang-Mills.

In this article we further develop the previous study.  
Random plane partitions are known 
\cite{Okounkov-Reshetikhin} to be described by two-dimensional 
free fermions (2$d$ conformal field theory). 
We compute the one-point functions of fermions 
in the statistical model and clarify their 
thermodynamic limits.
We start Section 2 with a brief review of the model. 
The interpretation as $q$-deformed random partitions 
is emphasized. In Section 3 we introduce one-point functions 
of fermions which are located at the main diagonal 
of the plane partition. They turn out to have a statistical 
interpretation.  
They are expressed as the statistical sums 
of suitable wave functions associated with partitions. 
Asymptotics of these wave functions, which are relevant  
to obtain the thermodynamic limit of the one-point functions, 
are computed.

As the $q$-deformed random partitions, 
the model can be specified by the Boltzmann weights 
attached to partitions rather than plane partitions. 
For a very large partition, the Boltzmann weight is 
measured by the energy function. 
In Section 4 and Appendix A, we compute the energy function 
in a way different from \cite{Nekrasov-Okounkov}.  
Our computations are based on the regularized density of the 
Maya diagram. The comparison with the known description 
using the profile is presented in Appendix B.

The thermodynamic limits or the semi-classical approximations of 
the fermion one-point functions at the main diagonal 
are proved in Section 5 to have 
a universal form in the sense that it is irrespective of detail 
of the minimizers of the energy function.  
They are given by the WKB type wave functions. 
In particular, at the four-dimensional limit, 
these semi-classical one-point functions are shown to be living 
on the Seiberg-Witten curve of the gauge theory.  
They are the WKB type wave functions 
whose classical action is precisely 
the integral of $dS_{s.w}$ on the curve, 
where $dS_{s.w}$ is the Seiberg-Witten differential.

The diagonal slices of plane partitions are labelled 
by the discretized time $m$. The main diagonal slice is 
at $m=0$. The discretized time becomes a continuous time $t$ 
at the thermodynamic limit. 
It is identified with a coordinate of the limit shape 
\cite{Okounkov-Reshetikhin} of random plane partition. 
In Section 6 we consider the $U(1)$ theory and 
compute the semi-classical wave functions of fermions 
which are located away from the main diagonal. 
We then argue the role of the parameter $t$ in gauge theories.

\section{A model of random plane partitions}

A plane partition $\pi$ is an array of non-negative integers 
\begin{eqnarray}
\begin{array}{cccc}
\pi_{11} & \pi_{12} & \pi_{13} & \cdots \\
\pi_{21} & \pi_{22} & \pi_{23} & \cdots \\
\pi_{31} & \pi_{32} & \pi_{33} & \cdots \\
\vdots & \vdots & \vdots & ~
\end{array}
\label{pi}
\end{eqnarray}
satisfying 
$\pi_{ij}\geq \pi_{i+1 j}$ and 
$\pi_{ij}\geq \pi_{i j+1}$ for 
all $i,j \geq 1$. 
Plane partitions are identified 
with the three-dimensional Young diagrams 
as depicted in Figure 1-(a). 
The three-dimensional diagram $\pi$ 
is a set of unit cubes such that $\pi_{ij}$ cubes 
are stacked vertically on each $(i,j)$-element of $\pi$. 
The size of $\pi$ is $|\pi| \equiv \sum_{i,j \geq 1}\pi_{ij}$, 
which is the total number of cubes of the diagram. 
Each diagonal slice of $\pi$ becomes a partition, 
that is, a sequence of weakly decreasing non-negative integers. 
See Figure 1-(b).  
Let $\pi(m)$ be a partition along the $m$-th diagonal slice. 
\begin{eqnarray}
\pi(m)= 
\left\{
\begin{array}{cl}
(\pi_{1~ m+1},\pi_{2~ m+2},\pi_{3~ m+3},\cdots) & 
~~\mbox{for}~m \geq 0 \\[1.5mm]
(\pi_{-m+1~ 1},\pi_{-m+2~ 2},\pi_{-m+3~ 3},\cdots) & 
~~\mbox{for}~m \leq -1.\\[1.5mm]
\end{array}
\right.
\end{eqnarray}
In particular 
$\pi(0)=(\pi_{11},\pi_{22},\pi_{33},\cdots)$ will be called 
the main diagonal partition. 
The series of partitions $\pi(m)$ satisfies the condition
\begin{eqnarray}
\cdots \prec \pi(-2) \prec \pi(-1) \prec 
\pi(0) \succ \pi(1) \succ \pi(2) \succ \cdots,
\end{eqnarray}
where $\mu \succ \nu$ means the following interlace relation 
between two partitions 
$\mu=(\mu_1 \geq \mu_2 \geq \cdots \geq 0)$ 
and 
$\nu=(\nu_1 \geq \nu_2 \geq \cdots \geq 0)$ 
\begin{eqnarray}
\mu \succ \nu ~~~
\Longleftrightarrow ~~~
\mu_1 \geq \nu_1 \geq \mu_2 \geq \nu_2 
\geq \mu_3 \geq \cdots.
\end{eqnarray} 
We have $|\pi|=\sum_{m=-\infty}^{+\infty}|\pi(m)|$,  
where the size of a partition $\mu$ is also denoted by  
$|\mu| \equiv \sum_{i \geq 1}\mu_i$. 
\begin{figure}[bt]
\begin{center}
\includegraphics[scale=0.7]{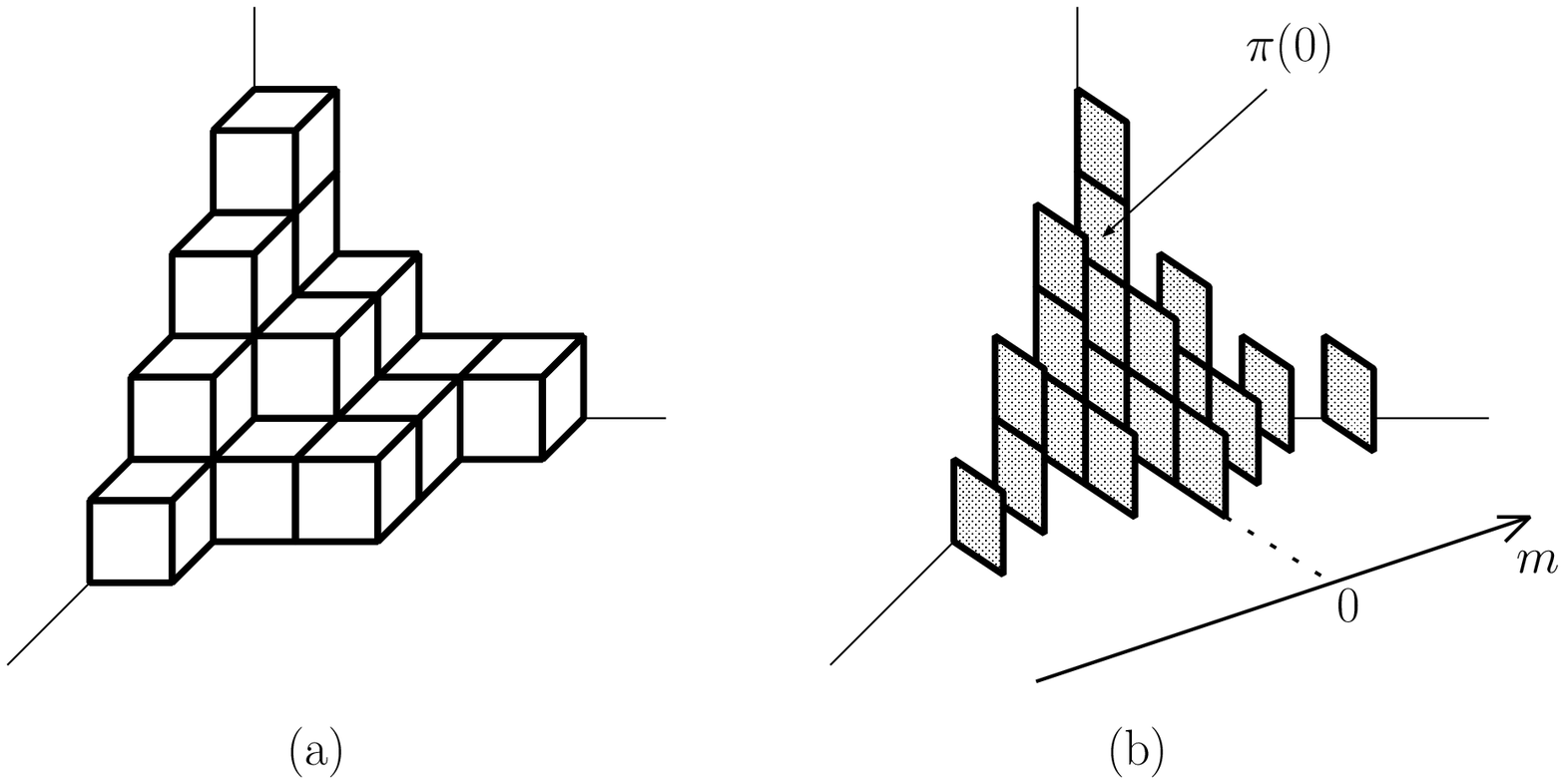}
\mycaption{1}{
The three-dimensional Young diagram (a) and 
the corresponding sequence of partitions 
(the two-dimensional Young diagrams) (b).}
\end{center}\label{3d Young diagram}
\end{figure}

A model of random plane partitions 
which describes five-dimensional ${\cal N}=1$ 
supersymmetric $SU(N)$ gauge theory is discussed 
in \cite{MNTT}. 
The partition function of the model is given by  
\begin{eqnarray}
Z^{SU(N)}_{q}
\equiv 
\sum_{\pi}\, q^{|\pi|}\,Q^{|\pi(0)|}\,e^{V(\pi(0))}. 
\label{Z SU(N)}
\end{eqnarray}
The Boltzmann weight consists of three parts. 
The first contribution comes from the energy of a plane 
partition $\pi$.  
The second contribution is a chemical potential 
for the main diagonal partition $\pi(0)$, and 
the third comes from the $N$-periodic potential 
\cite{Nekrasov-Okounkov} for the main diagonal partition. 
Let $\xi_r \in \mathbb{R}$ ($1\leq r \leq N$) be 
such that $\sum_{r=1}^N \xi_r=0$, and 
define the $N$-periodic potential $v$ on $\mathbb{Z}$ 
by $v(k)=\xi_{k+1 \mbox{ \scriptsize{mod} } N}$. 
The potential $V$ for a partition $\mu$ is formally 
given \footnote{See Appendix A for more information.} by
\begin{eqnarray}
V(\mu)=
\sum_{i=1}^{+\infty} 
v(\mu_i-i).
\label{N-periodic potential}
\end{eqnarray} 
To contact with the $SU(N)$ gauge theory 
we also need to identify \cite{MNTT} 
the indeterminates $q$ and $Q$ 
with the following field theory variables. 
\begin{eqnarray}
q=e^{-\frac{2R}{N}\hbar},~~~
Q=(2R\Lambda)^2, 
\label{q and Q}
\end{eqnarray}
where $R$ is the radius of $S^1$ in the fifth dimension, 
and $\Lambda$ is the lambda parameter of the underlying 
four-dimensional field theory.

\subsubsection*{Transfer matrix approach}

The transfer matrix approach \cite{Okounkov-Reshetikhin} 
allows us to express the random plane partitions 
(\ref{Z SU(N)}) in terms of two-dimensional conformal 
field theory (2$d$ free fermion system).

It is well known that partitions 
are realized as states of 2$d$ free fermions 
by using the Maya diagrams. 
Let 
$\psi(z)=\sum_{k \in \mathbb{Z}}
\psi_k z^{-k-1}$
and 
$\psi^*(z)=\sum_{k \in \mathbb{Z}}
\psi^*_k z^{-k}$
be complex fermions with the anti-commutation relations 
\begin{eqnarray}  
\left\{ \psi_k,\psi^*_l \right\}
=\delta_{k+l,0},~~~~
\left\{ \psi_k,\psi_l \right\}
=\left\{ \psi_k^*,\psi_l^* \right\}
=0. 
\end{eqnarray}
Let 
$\mu=(\mu_1, \mu_2, \cdots)$ be a partition. 
The Maya diagram $\mu$ is 
a series of the strictly decreasing integers 
$x_{i}(\mu) 
\equiv -i+\mu_i$, 
where $i \in \mathbb{Z}_{\geq 1}$. 
The correspondence with the Young diagram 
is depicted in Figure 2.
By using the Maya diagram 
the partition can be mapped to 
the following fermion state 
\begin{eqnarray}
|\mu;n \rangle 
=
\psi_{-x_1(\mu)-1-n}\psi_{-x_2(\mu)-1-n}
\cdots \psi_{-x_{l(\mu)}(\mu)-1-n}
\psi^*_{-l(\mu)+1+n}\psi^*_{-l(\mu)+2+n}
\cdots \psi^*_{n} 
|\emptyset;\,n \rangle, 
\label{|mu;n>}
\end{eqnarray}
where $l(\mu)$ is the length of $\mu$, that is, 
the number of the non-zero $\mu_i$. 
In (\ref{|mu;n>}) the state $|\emptyset;\,n\rangle$ is 
the ground state of the charge $n$ sector. 
It is defined by the conditions 
\begin{eqnarray}
&&
\psi_{k}|\emptyset;\,n \rangle=0~~~\mbox{for}~k \geq -n,
\nonumber \\
&&
\psi_{k}^*|\emptyset;\,n \rangle=0~~~\mbox{for}~k \geq n+1.
\label{|phi;n>}
\end{eqnarray}
We mainly consider the $n=0$ sector in the below. 
\begin{figure}[hbt]
\begin{center}
\includegraphics[scale=0.5]{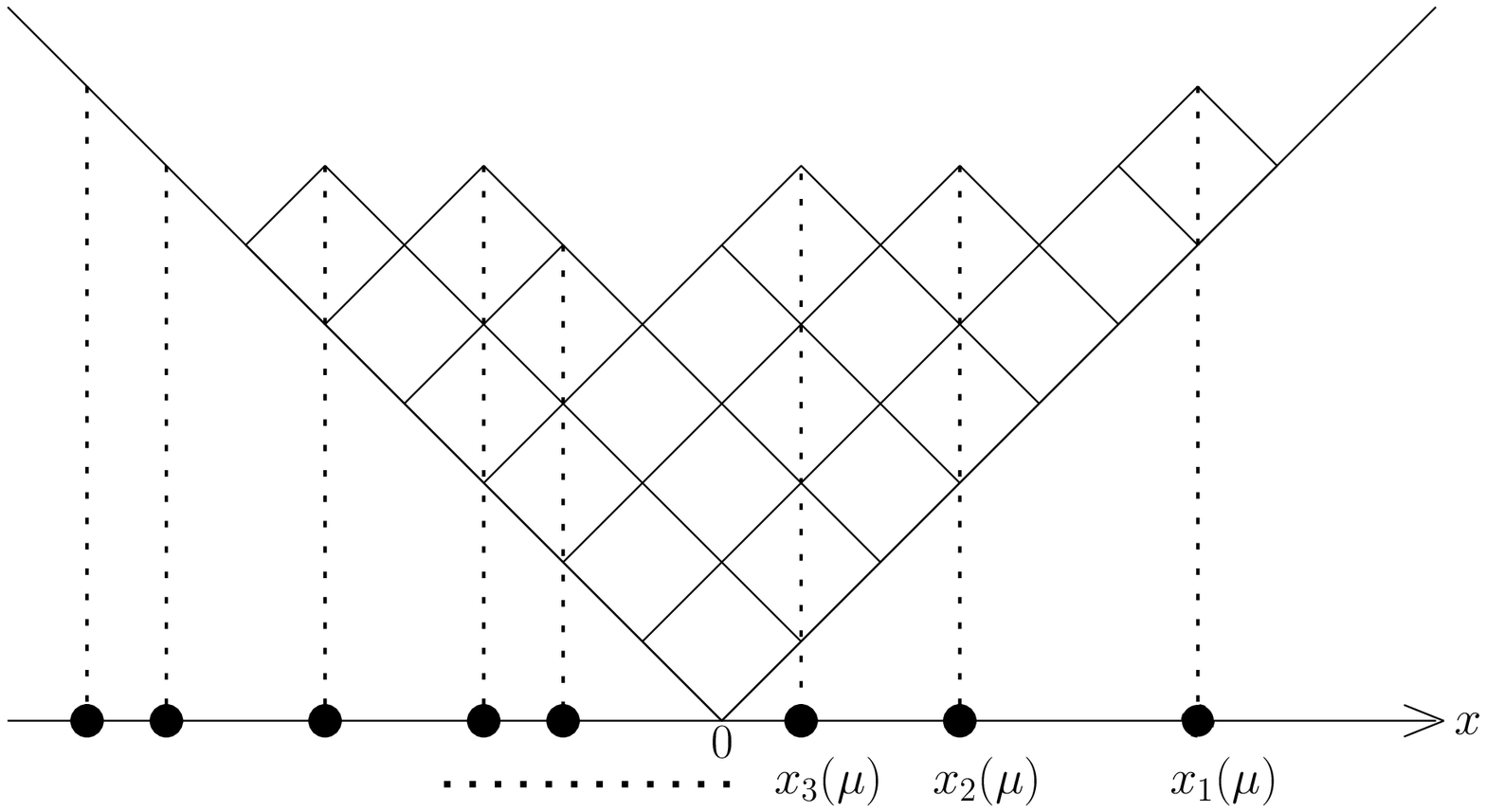}
\mycaption{2}{The correspondence between 
the Maya diagram and 
the Young diagram of $\mu=(7,5,4,2,2,1)$.
Elements of the Maya diagram 
are denoted by $\bullet$.}
\end{center}\label{fig1}
\end{figure}

The basic ingredient of the transfer matrix approach 
is the following evolution operator at a discretized time 
$m \in \mathbb{Z}$.
\begin{eqnarray}
\Gamma(m)
&\equiv& 
\left\{
\begin{array}{cl}
\exp\Bigl(
\displaystyle{\sum_{k=1}^{+\infty}}
\frac{1}{k}q^{k(m+\frac{1}{2})}J_{-k}\Bigr)
& 
~~~\mbox{for}~m \geq 0 \\[1.5mm]
\exp\Bigl(
\displaystyle{\sum_{k=1}^{+\infty}}
\frac{1}{k}q^{-k(m+\frac{1}{2})}J_{k}\Bigr)
& 
~~~\mbox{for}~m \leq -1,  \\[1.5mm]
\end{array}
\right.
\label{gamma m}
\end{eqnarray}
where $J_{\pm k}$ are the modes of the standard $U(1)$ current 
\begin{eqnarray}
:\psi \psi^*:(z)=\sum_{n}J_nz^{-n-1}.
\end{eqnarray}
Implications of the above operators 
in random plane partitions can be 
understood from their matrix elements: 
For $m \geq 0$, 
\begin{eqnarray}
\langle \mu;\,0|
\Gamma(m)
|\nu;\,0 \rangle 
=
\left\{
\begin{array}{cl}
q^{(m+\frac{1}{2})(|\mu|-|\nu|)}  &
\mu \succ \nu    \\[1.5mm]
0 & 
\mbox{otherwise}, \\[1.5mm]
\end{array}
\right.
\label{gamma m on mu 1}
\end{eqnarray}
and for $m \leq -1$,  
\begin{eqnarray}
\langle \mu;\,0|
\Gamma(m)
|\nu;\,0 \rangle 
=
\left\{
\begin{array}{cl}
q^{(m+\frac{1}{2})(|\mu|-|\nu|)}  &
\mu \prec \nu    \\[1.5mm]
0 & 
\mbox{otherwise}. \\[1.5mm]
\end{array}
\right.
\label{gamma m on mu 2}
\end{eqnarray}
It follows from (\ref{gamma m on mu 1}) 
and (\ref{gamma m on mu 2}) that the partition 
function is expressed as 
\begin{eqnarray}
Z^{SU(N)}_q=
\langle \emptyset;\,0|
\Biggl\{ 
\prod_{m \leq -1} \Gamma(m)
\Biggr\}
Q^{L_0}
\exp 
\Biggl(
\sum_{r=1}^{N}\xi_r J_{0}^{(r)}
\Biggr)
\Biggl\{
\prod_{m \geq 0}  \Gamma(m)
\Biggr\}
| \emptyset;\, 0\rangle.
\label{Z SU(N) transfer matrix}
\end{eqnarray}
Here, 
$J_0^{(r)}$ are the zero modes of 
the $U(1)$ currents of $N$-component fermions 
$\psi^{(r)}$ and $\psi^{(r)*}$ which are 
realized \cite{Miwa-Jimbo} by folding $\psi$ and $\psi^{*}$.

\subsubsection*{$q$-deformed random partitions}

We can interpret the random plane partitions 
(\ref{Z SU(N)}) as a model of random partitions. 
It is identified with a $q$-deformation of the 
random partitions \cite{Nekrasov-Okounkov}. 
(The deformation itself differs from \cite{Nekrasov-Okounkov}.) 
To see this, 
we rewrite the partition function 
by using the Schur functions. 
An insertion of the unity 
$1=\sum_{\mu}|\mu;\,0 \rangle \langle \mu ;\,0|$ 
factorizes (\ref{Z SU(N) transfer matrix}) into 
\begin{eqnarray}
Z_q^{SU(N)}&=&
\sum_{\mu}
Q^{|\mu|}e^{V(\mu)}
\langle \emptyset;\,0|
\prod_{m \leq -1}
\Gamma(m)
|\mu;\, 0\rangle 
\langle \mu;\,0| 
\prod_{m \geq 0}
\Gamma(m)
|\emptyset;\,0\rangle.  
\label{factorized Z SU(N) transfer matrix}
\end{eqnarray}
The matrix elements in the above turn to be 
\begin{eqnarray}
\langle \emptyset ;\,0|
\prod_{m \leq -1}\Gamma(m) 
| \mu ;\, 0 \rangle 
&=&
\langle \emptyset ;\,0 | 
\prod_{k=1}^{+\infty}
\exp \left( 
\frac{1}{k}
\sum_{i=1}^{+\infty}
q^{k(i-\frac{1}{2})}
J_k \right) 
| \mu; \,0 \rangle 
\nonumber \\[1.5mm]
&=&
s_{\mu}(q^{\frac{1}{2}},q^{\frac{3}{2}},\cdots), 
\label{schur mu 1} \\[1.5mm]
\langle \mu ;\,0|
\prod_{m \geq 0}\Gamma(m) 
| \emptyset ;\, 0 \rangle 
&=&
\langle \mu;0 | 
\prod_{k=1}^{+\infty}
\exp \left( 
\frac{1}{k}
\sum_{i=1}^{+\infty}
q^{k(i-\frac{1}{2})}
J_{-k} \right) 
| \emptyset ;\, 0 \rangle 
\nonumber \\[1.5mm]
&=&
s_{\mu}(q^{\frac{1}{2}},q^{\frac{3}{2}},\cdots), 
\label{schur mu 2}
\end{eqnarray} 
where 
$s_{\mu}(q^{\frac{1}{2}},q^{\frac{3}{2}},\cdots)$ 
is the Schur function 
$s_{\mu}(x_1,x_2,\cdots)$ specialized at 
$x_i=q^{i-\frac{1}{2}}$ ($i \geq 1$). 
Therefore we obtain 
\begin{eqnarray}
Z_q^{SU(N)}
&=&
\sum_{\mu}\,Q^{|\mu|}e^{V(\mu)}
s_{\mu}(q^{-\rho})^2, 
\label{Z SU(N) schur}
\end{eqnarray}
where the multiple index 
$\rho \equiv (-\frac{1}{2},-\frac{3}{2}, 
\cdots, -i+\frac{1}{2},\cdots)$ is used. 
The expression (\ref{Z SU(N) schur}) allows us 
to interpret (\ref{Z SU(N)}) 
as a model of $q$-deformed random partitions. 
It is also clear from  
(\ref{factorized Z SU(N) transfer matrix}) 
that partitions $\mu$ are the main diagonal 
partitions $\pi(0)$.

The four-dimensional limit of the model  
is obtained by letting $R \rightarrow 0$ 
under the identification (\ref{q and Q}). 
To take the limit the following 
product formula of the Schur function 
\cite{Macdonald} becomes useful. 
\begin{eqnarray}
s_{\mu}(q^{-\rho})
&=&
q^{n(\mu)+\frac{1}{2}|\mu|}
\prod_{(i,j)\in \mu}
\frac{1}
{\left(1-q^{h(i,j)}\right)}, 
\label{product formula b}
\end{eqnarray}
where $h(i,j)$ is the hook length of the box $(i,j)$ 
in the Young diagram,  
and $n(\mu)\equiv \sum_{i \geq 1}(i-1)\mu_i$. 
By using (\ref{product formula b}) 
we can see from (\ref{Z SU(N) schur}) that 
\begin{eqnarray}
\lim_{R \rightarrow 0}Z_q^{SU(N)}
=
\sum_{\mu}
\left(\frac{N\Lambda}{\hbar}\right)^{2|\mu|}
e^{V(\mu)}
\left(\frac{dim \mu}{|\mu|!}\right)^2. 
\label{4d U(1) random partitions}
\end{eqnarray}
This is the model of random partitions \cite{Nekrasov-Okounkov}
which describes four-dimensional 
${\cal N}=2$ supersymmetric $SU(N)$ Yang-Mills 
at the thermodynamic limit.

Thermodynamic limit of the model is achieved  
by letting $\hbar \rightarrow 0$. To see this, 
we consider the $U(1)$ theory or the $SU(N)$ theory with 
the periodic potential turned off. 
In such a situation the partition function can be computed 
by using the standard technique of 
2$d$ conformal field theory.  
It becomes 
\begin{eqnarray}
Z_q^{U(1)}=\prod_{n=1}^{+\infty}
(1-Qq^n)^{-n}.
\label{MacMahon}
\end{eqnarray}
The mean values of $|\pi|$ and $|\pi(0)|$ 
are respectively given by $q\frac{\partial}{\partial q}\ln Z$ 
and $Q\frac{\partial}{\partial Q}\ln Z$. 
It follows from (\ref{MacMahon}) that 
they behave 
$\left \langle |\pi| \right \rangle =O(\hbar^{-3})$ and 
$\left \langle |\pi(0)| \right \rangle =O(\hbar^{-2})$ 
as $\hbar \rightarrow 0$ ($q \rightarrow 1$). 
Therefore a typical plane partition $\pi$ 
near the limit $\hbar \rightarrow 0$ is a plane partition 
of order $\hbar^{-3}$, 
and its main diagonal partition $\pi(0)$ or $\mu$
becomes a partition of order $\hbar^{-2}$.

\newpage
\section{One-point functions at the main diagonal}

In the present section and section 5  
we investigate one-point functions of the free fermions 
located at the main diagonal of the plane partition $\pi$.  
From the transfer matrix viewpoint they are given by 
\begin{eqnarray}
\Psi_q^*(z) 
&\equiv& 
\frac{1}{Z^{SU(N)}_q}
\langle \emptyset;\, -1 |
\Biggl\{
\prod_{m \leq -1} \Gamma(m)
\Biggr\}
\,
\psi^*(z)
\,
q^{\frac{1}{2}L_0}
Q^{L_0}
\,
e^{\sum_{r=1}^{N}\xi_r J_{0}^{(r)}}
\Biggl\{
\prod_{m \geq 0}\Gamma(m)
\Biggr\}
| \emptyset;\, 0 \rangle,
\nonumber \\*
&&
\label{Psi(lambda)star}\\
\Psi_q(z) 
&\equiv& 
\frac{1}{Z^{SU(N)}_q}
\langle \emptyset;\, +1 |
\Biggl\{
\prod_{m \leq -1} \Gamma(m)
\Biggr\}
\,
\psi(z)
\,
q^{-\frac{1}{2}L_0}
\,
Q^{L_0}
\,
e^{\sum_{r=1}^{N}\xi_r J_{0}^{(r)}}
\Biggl\{
\prod_{m \geq 0}\Gamma(m)
\Biggr\}
| \emptyset ;\, 0 \rangle, 
\nonumber \\*
&&
\label{Psi(lambda)}
\end{eqnarray}
where $q^{\pm \frac{L_0}{2}}$ are inserted 
for the later convenience of normalization.

The goal of our discussions is to clarify
the thermodynamic limits of the above one-point functions. 
These are treated in section 5 
by using the results of this section.

\subsubsection*{Statistical interpretation of 
fermion one-point functions}

We regard the random plane partitions (\ref{Z SU(N)}) 
as the $q$-deformed random partitions (\ref{Z SU(N) schur}). 
The Boltzmann weight for a partition $\mu$ is given by 
$Q^{|\mu|}e^{V(\mu)}s_{\mu}(q^{-\rho})^2$. 
The above one-point functions admit to have a statistical 
interpretation. As we confirm subsequently, they can be
expressed as the following statistical sums. 
\begin{eqnarray}
\Psi_q^*(z)
&=&
\frac{1}{Z^{SU(N)}_q}
\sum_{\mu}\,
\chi_q^*(z \mid \mu)\,\,
Q^{|\mu|}
e^{V(\mu)}
s_{\mu}(q^{-\rho})^2,
\label{Psi(lambda)star 2} 
\\
\Psi_q(z)
&=&
\frac{1}{Z_q^{SU(N)}}
\sum_{\mu}\, 
\chi_q(z \mid \mu)\,\, 
Q^{|\mu|}
e^{V(\mu)}
s_{\mu}(q^{-\rho})^2,  
\label{Psi(lambda) 2} 
\end{eqnarray}
where $\chi_q^*(z\mid\mu)$ and 
$\chi_q(z\mid\mu)$ are thought as wave functions 
associated with each partition $\mu$. 
They are defined by  
\begin{eqnarray}
\chi_q^*(z \mid \mu)
&\equiv& 
\sum_{i=1}^{+\infty}
z^{-x_i(\mu)-1}
(-)^{i-1}
q^{\frac{1}{2}|\mu|}
\frac{s_{(\mu:\, i)}(q^{-\rho})}{s_{\mu}(q^{-\rho})}, 
\label{chi star} 
\\
\chi_q(z \mid \mu)
&\equiv& 
\sum_{i=1}^{+\infty}
z^{-x_i(\widetilde{\mu})-1}
(-)^{i-1}
q^{-\frac{1}{2}|\mu|}
\frac{s_{(\widetilde{\mu}:\, i)}(q^{\rho})}
{s_{\widetilde{\mu}}(q^{\rho})}.  
\label{chi} 
\end{eqnarray}
Here $(\mu:i)$ is a partition obtained from 
$\mu$ for each $i \in \mathbb{Z}_{\geq 1}$. 
It is described 
in terms of the Maya diagram as follows.    
\begin{eqnarray}
x_{k}(\mu:i)=
\left \{
\begin{array}{cl}
x_k(\mu)+1 &~~\mbox{for}~~ 1 \leq k <i \\[1.5mm]
x_{k+1}(\mu)+1 &~~\mbox{for}~~ k \geq i. \\
\end{array}
\right.
\label{x(mu:i)}
\end{eqnarray}
We can see from the above that 
$|(\mu:i)|=|\mu|-x_i(\mu)-1$. 
The partition conjugate to $\mu$ is denoted 
by $\widetilde{\mu}$ in (\ref{chi}).
The corresponding Young diagram is 
obtained by flipping the Young diagram $\mu$ 
over its main diagonal. 
One can find a relation between 
$\chi_q(z \mid \mu)$ and $\chi_q^*(z \mid \mu)$.
By comparing (\ref{chi star}) with (\ref{chi}) 
we obtain 
\begin{eqnarray}
\chi_q(z \mid \mu)=
\chi_{q^{-1}}^*
(z \mid \widetilde{\mu}). 
\label{chi vs chi star}
\end{eqnarray}


~

Let us verify the descriptions 
(\ref{Psi(lambda)star 2}) and (\ref{Psi(lambda) 2}). 
We first consider the one-point function $\Psi_q^*(z)$. 
By the insertion of the unity 
$1=\sum_{\mu}|\mu;\,0 \rangle \langle \mu ;\,0|$ 
the following factorization of (\ref{Psi(lambda)star}) 
is obtained. 
\begin{eqnarray}
\frac{1}{Z_q^{SU(N)}}
\sum_{\mu}\,
Q^{|\mu|}
e^{V(\mu)}
q^{\frac{1}{2}|\mu|}\,
\langle \emptyset ;\,-1 | 
\Biggl\{
\prod_{m \leq -1}\Gamma(m)
\Biggr\}
\psi^*(z)\,
|\mu;\, 0 \rangle 
\langle \mu;\, 0 |
\prod_{m \geq 0}\Gamma(m)
|\emptyset;\,0 \rangle.   
\label{Psi star proof}
\end{eqnarray}
This can be written down by using the Schur functions. 
The second component is just (\ref{schur mu 2}). 
In order to express the first component in terms of 
the Schur functions, it is convenient to 
describe the action of 
$\psi^*(z)$ on the partition $\mu$ explicitly. 
By using (\ref{|mu;n>}) we can see 
that the actions of $\psi^*_k$ on the partition become
\begin{eqnarray}
\psi^*_k\,\,
|\mu;\,0 \rangle 
=
\left\{
\begin{array}{cl}
(-)^{i-1}|(\mu:i);\,-1\rangle &~~~ 
\mbox{if}~~k=x_{i}(\mu)+1 \\[1.5mm]
0 & ~~~\mbox{otherwise},\\
\end{array}
\right.
\label{psi star mode on mu}
\end{eqnarray}
where $(\mu:i)$ is the partition given 
in (\ref{x(mu:i)}). 
The above actions are translated to  
\begin{eqnarray}
\psi^*(z)\,|\mu;\,0\rangle 
=
\sum_{i=1}^{+\infty}
z^{-x_i(\mu)-1}
(-)^{i-1}
|(\mu:i);\,-1\rangle. 
\label{psi star on mu}
\end{eqnarray}
By using (\ref{psi star on mu}) and also 
taking (\ref{schur mu 1}) into account, 
the first component of (\ref{Psi star proof}) becomes  
\begin{eqnarray}
\langle \emptyset ;\,-1 | 
\Biggl\{
\prod_{m \leq -1}\Gamma(m)
\Biggr\}
\psi^*(z)\,
|\mu;\, 0 \rangle 
=
\sum_{i=1}^{+\infty}
z^{-x_i(\mu)-1}
(-)^{i-1}
s_{(\mu:\,i)}(q^{-\rho}). 
\label{1st factor}
\end{eqnarray}
The description (\ref{Psi(lambda)star 2}) is obtained 
from (\ref{Psi(lambda)star}) by plugging  
(\ref{1st factor}) and (\ref{schur mu 2}) 
into (\ref{Psi star proof}) and then 
arranging them noting the definition (\ref{chi star}).

Nextly we consider the case of $\Psi_q(z)$.  
The description (\ref{Psi(lambda) 2}) can be obtained 
from (\ref{Psi(lambda)}) by taking the same steps as above. 
We first factorize (\ref{Psi(lambda)}) into 
\begin{eqnarray}
\frac{1}{Z_q^{SU(N)}}
\sum_{\mu}
Q^{|\mu|}
e^{V(\mu)}
q^{-\frac{|\mu|}{2}}
\langle \emptyset;\,+1 | 
\Biggl\{
\prod_{m \leq -1}\Gamma(m)
\Biggr\}
\psi(z)\,
|\mu;\, 0 \rangle 
\langle \mu;\, 0 |
\prod_{m \geq 0}\Gamma(m)
|\emptyset ;\,0 \rangle.   
\label{Psi proof}
\end{eqnarray}
To write down the first component in terms of 
the Schur functions we compute the action of 
$\psi(z)$ on the partition $\mu$. 
The actions of $\psi_k$ on the partition  
can be read as 
\begin{eqnarray}
\psi_k
|\mu;\,0 \rangle 
=
\left\{
\begin{array}{cl}
(-)^{i+x_i(\widetilde{\mu})}
|\widetilde{(\widetilde{\mu}:i)};\,+1\rangle & 
~~~\mbox{if}~~
k=x_{i}(\widetilde{\mu}) \\[1.5mm]
0 & 
~~~\mbox{otherwise}\,. 
\end{array}
\right.
\label{psi mode on mu}
\end{eqnarray}
These are translated to 
\begin{eqnarray}
\psi(z)\,|\mu;\,0\rangle =
\sum_{i=1}^{+\infty}
z^{-x_i(\widetilde{\mu})-1}
(-)^{i+x_i(\widetilde{\mu})}
|\widetilde{(\widetilde{\mu}:i)};\,+1 \rangle. 
\label{psi on mu}
\end{eqnarray}
Therefore the first component of (\ref{Psi proof}) becomes  
\begin{eqnarray}
\langle \emptyset ;\, +1 |
\Biggl\{
\prod_{m \leq -1} \Gamma(m)
\Biggr\}
\psi(z)\,
| \mu;\, 0 \rangle 
=\sum_{i=1}^{+\infty}
z^{-x_i(\widetilde{\mu})-1}
(-)^{i+x_i(\widetilde{\mu})}
s_{\widetilde{(\widetilde{\mu}:\,i)}}
(q^{-\rho}). 
\label{1st factor 2}
\end{eqnarray}
The description (\ref{Psi(lambda) 2}) is obtained 
from (\ref{Psi(lambda)}) by 
plugging (\ref{1st factor 2}) and (\ref{schur mu 2}) 
into (\ref{Psi proof}) and making use of the identity 
$s_{\widetilde{\mu}}(q^{-\rho})=
(-)^{|\mu|}s_{\mu}(q^{\rho})$.

\subsubsection*{Large $|\mu|$ behaviors of 
wave functions associated with partitions}

Near the thermodynamic limit, 
which is achieved by letting 
$\hbar \rightarrow 0$ in the model, 
very large partitions dominate. 
Their size becomes of order $\hbar^{-2}$.
We will clarify the asymptotics of 
$\chi_q^*(z\mid \mu)$ and $\chi_q(z\mid \mu)$ 
when $\mu$ is such a large partition. 
The Maya diagram of such a partition 
can be thought as a quantity of order $\hbar^{-1}$. 
For the description we conveniently rescale $x_i(\mu)$. 
Let us introduce the order $\hbar^0$ quantities 
$u$ and $s$ by the following rescalings. 
\begin{eqnarray}
x=\frac{N}{\hbar}u,~~~~~
i=\frac{N}{\hbar}s, 
\label{u and s}
\end{eqnarray}
where $u \in \mathbb{R}$ and $s \in \mathbb{R}_{\geq 0}$. 
Correspondingly, 
the Maya diagram is scaled to a function $u(s\mid \mu)$ as
\begin{eqnarray}
x_{i}(\mu)=\frac{N}{\hbar}u(s\mid \mu)+O(\hbar^0).
\label{u(s)}
\end{eqnarray} 
Notice that the Maya diagram is a strictly decreasing series 
satisfying the conditions that 
$x_{i}(\mu) \geq -i$ for $\forall i \in \mathbb{Z}_{\geq 1}$ 
and that $x_{i}(\mu)=-i$ for sufficiently large $i$. 
Therefore, the function $u(s\mid \mu)$ is monotonically decreasing,  
and satisfies the conditions that 
$u(s \mid\mu) \geq -s$ for $\forall s \in 
\mathbb{R}_{\geq 0}$ and that $u(s\mid \mu) \rightarrow -s$ 
as $s \rightarrow +\infty$.

The large $|\mu|$ behaviors of the wave functions 
can be computed from (\ref{chi star}) and (\ref{chi}) 
by scaling $\mu$ as (\ref{u(s)}). As the result, 
they turn out to be  
\begin{eqnarray}
&&
\chi_q^*(z\mid \mu)
\,\approx\,
\frac{N}{\hbar}
\int_0^{+\infty}
d\underline{s}
~z^{-\frac{N}{\hbar}u(\underline{s}\mid\, \mu)}
\nonumber \\[1.5mm]
&&
~~~~~~~~~
\times 
\exp \left\{
-\frac{RN}{2\hbar}u(\underline{s}\mid \mu)^2
+\frac{N}{\hbar}
\int_{-\infty}^{+\infty}
dv \,
\frac{ds(v\mid \mu)}{dv}
\ln 
\left(
\frac{\sinh R(u(\underline{s}\mid\mu)-v)}{R\Lambda}
\right)
\right\}, 
\nonumber \\[1.5mm]
\label{asymptotic chi star} \\[1.5mm]
&&
\chi_q(z\mid \mu)
\,\approx\,
\frac{N}{\hbar}
\int_0^{+\infty}
d\underline{s}
~z^{-\frac{N}{\hbar}u(\underline{s}\mid\, \widetilde{\mu})}
\nonumber \\*[1.5mm]
&&
~~~~~~~~~
\times 
\exp \left\{
\frac{RN}{2\hbar}u(\underline{s}\mid \widetilde{\mu})^2
-\frac{N}{\hbar}
\int_{-\infty}^{+\infty}
dv\, 
\frac{ds(v\mid \mu)}{dv}
\ln 
\left(
\frac{\sinh R(-u(\underline{s}\mid\widetilde{\mu})-v)}{R\Lambda}
\right)
\right\}. 
\nonumber \\*[1.5mm]
\label{asymptotic chi}
\end{eqnarray}
In the above we have used the density function 
$\frac{d}{du}s(u\mid\mu)$. 
The function $s(u\mid\mu)$ is 
the inverse of $u(s\mid\mu)$, that is,  
$u(s(v|\,\mu)\,|\,\mu)=v$. 
It becomes non-negative and weakly decreasing over $\mathbb{R}$, 
satisfying the condition that 
$s(u\mid\mu) \geq -u$. 
The asymptotic behaviors are 
$s(u\mid\mu) \rightarrow -u$ as $u \rightarrow -\infty$ 
and
$s(u\mid\mu) \rightarrow 0$ as $u \rightarrow +\infty$. 
Therefore, 
the density function 
$\frac{d}{du}s(u\mid\mu)$ 
takes values in $[-1,0]$, 
and satisfies the asymptotic conditions 
that 
$\frac{d}{du}s(u\mid\mu)\rightarrow -1$ 
as 
$u \rightarrow -\infty$ and that 
$\frac{d}{du}s(u\mid\mu)\rightarrow 0$ 
as 
$u \rightarrow +\infty$. 
We should note 
the density function in the proper sense
\footnote{See (\ref{rho vs ds}) in Appendix A.} is 
$-\frac{d}{du}s(u\mid\mu)$ rather than 
$\frac{d}{du}s(u\mid\mu)$. 
However, in this article,  
$\frac{d}{du}s(u\mid\mu)$ is called the density function 
as well, unless it causes any confusion.

Before confirming the asymptotics 
(\ref{asymptotic chi star}) and (\ref{asymptotic chi}), 
we shall describe their four-dimensional limits.  
The limits are obtained from 
(\ref{asymptotic chi star}) and (\ref{asymptotic chi}) 
by letting $R \rightarrow 0$. 
They become  
\begin{eqnarray}
&&
\chi_{4d}^{*}(z\mid\mu)
\, \equiv \, 
\lim_{R \rightarrow 0}
\chi_q^*(z\mid \mu)
\nonumber \\*[1.5mm]
&&~~~~~~
\, \approx \,
\frac{N}{\hbar}
\int_0^{+\infty}
d\underline{s}
~z^{-\frac{N}{\hbar}u(\underline{s} \mid \, \mu)}
\exp \left\{
\frac{N}{\hbar}
\int_{-\infty}^{+\infty}
dv \,
\frac{ds(v\mid \mu)}{dv}
\ln 
\left(
\frac{u(\underline{s}\mid\mu)-v}{\Lambda}
\right)
\right\}, 
\nonumber \\* 
\label{4d asymptotic chi star} \\[1.5mm]
&&
\chi_{4d}(z\mid\mu)
\, \equiv \, 
\lim_{R \rightarrow 0}
\chi_q(z\mid \mu)
\nonumber \\*[1.5mm]
&&~~~~~~
\,\approx \, 
\frac{N}{\hbar}
\int_0^{+\infty}
d\underline{s}
~z^{-\frac{N}{\hbar}u(\underline{s}\mid\, \widetilde{\mu})}
\exp \left\{
-\frac{N}{\hbar}
\int_{-\infty}^{+\infty}
dv \,
\frac{ds(v\mid \mu)}{dv}
\ln 
\left(
\frac{-u(\underline{s}\mid\widetilde{\mu})-v}{\Lambda}
\right)
\right\}. 
\nonumber \\*[1.5mm] 
&&
\label{4d asymptotic chi}
\end{eqnarray}


~

Let us derive the asymptotics 
(\ref{asymptotic chi star}) and (\ref{asymptotic chi}). 
Consider the case of $\chi^*_q(z\mid\mu)$. 
The following product formula of 
the (specialized) Schur function 
\cite{Macdonald} becomes useful to obtain the asymptotics.
\begin{eqnarray}
s_{\mu}(q^{-\rho})=
q^{-\frac{1}{4}\kappa(\mu)}
\prod_{1 \leq i<j < +\infty}
\left\{
\displaystyle{
\frac{q^{\frac{1}{2}(x_j(\mu)-x_i(\mu))}-
          q^{\frac{1}{2}(x_i(\mu)-x_j(\mu))}}
     {q^{\frac{1}{2}(i-j)}-q^{\frac{1}{2}(j-i)}}}
\right\}, 
\label{product formula a} 
\end{eqnarray}
where $\kappa(\mu)\equiv 2\sum_{(i,j)\in \mu}(j-i)$.

We translate the Schur functions in (\ref{chi star}) 
into the infinite products by using the above formula. 
$x_k(\mu:i)$ which appear in the product representation 
of $s_{(\mu:\,i)}(q^{-\rho})$ can be converted to 
$x_k(\mu)$ by using (\ref{x(mu:i)}). 
It turns out that almost all the products cancel between 
$s_{\mu}(q^{-\rho})$ and $s_{(\mu:\,i)}(q^{-\rho})$.  
Finally the ratio of the Schur functions 
in (\ref{chi star}) becomes as follows;   
\begin{eqnarray}
\frac{s_{(\mu:\,i)}(q^{-\rho})}
{s_{\mu}(q^{-\rho})}
&=&
q^{-\frac{1}{2}|\mu|+
\frac{1}{4}(x_i(\mu)+1)(x_i(\mu)+2)}
\displaystyle{
  \prod_{1 \leq j \leq i-1}
  \left( 
  q^{\frac{1}{2}(j-i)}-q^{\frac{1}{2}(i-j)}
  \right)^{-1}} 
\nonumber \\*[1.5mm]
&&
\times 
\displaystyle{
    \prod_{j \neq i}^{+\infty}
    \left \{
       \frac{q^{\frac{1}{2}(x_j(\mu)-x_i(\mu))}
             -q^{\frac{1}{2}(x_i(\mu)-x_j(\mu))}}
            {q^{\frac{1}{2}(i-j)}-q^{\frac{1}{2}(j-i)}}
             \right \}^{-1}}. 
\label{maeda formula}
\end{eqnarray}
This gives the following expression 
of $\chi_q^*(z\mid\mu)$; 
\begin{eqnarray}
\chi_q^*(z \mid\mu)=
\sum_{i=1}^{+\infty}\, 
z^{-x_i(\mu)-1}\,
q^{\frac{1}{4}(x_i(\mu)+1)(x_i(\mu)+2)}\,
\exp \Bigl\{\Phi_q(i\mid\mu)\Bigr\}, 
\label{chi star 2} 
\end{eqnarray}
where $\Phi_q(i\mid\mu)$ is defined by 
\begin{eqnarray}
\exp 
\Bigl\{\Phi_q(i\mid\mu) \Bigr\}
\equiv
\displaystyle{ 
\prod_{j=1}^{i-1}
\left( 
  q^{\frac{1}{2}(i-j)}-q^{\frac{1}{2}(j-i)}
\right)^{-1}}
\, 
\displaystyle{ 
\prod_{j \neq i}^{+\infty}
\left \{
   \frac{q^{\frac{1}{2}(x_j(\mu)-x_i(\mu))}
         -q^{\frac{1}{2}(x_i(\mu)-x_j(\mu))}}
        {q^{\frac{1}{2}(i-j)}-q^{\frac{1}{2}(j-i)}}
\right \}^{-1}}. 
\label{Phi}
\end{eqnarray}
The similar expression of $\chi_q(z\mid\mu)$ 
is obtainable from (\ref{chi star 2}) by using 
(\ref{chi vs chi star}).

The large $|\mu|$ behavior of $\chi_q^*(z\mid\mu)$ 
can be extracted from (\ref{chi star 2}) 
by scaling $x_i(\mu)$ as prescribed in (\ref{u(s)}). 
Taking account of $q=e^{-2R\hbar/N}$ 
we first observe   
\begin{eqnarray}
q^{\frac{1}{4}(x_i(\mu)+1)(x_i(\mu)+2)}
=\exp
\left\{
-\frac{RN}{2\hbar}u(\underline{s}\mid\mu)^2
+O(\hbar^0) 
\right\}, 
\label{asymptotic K}
\end{eqnarray}
where we put $i=\frac{N}{\hbar}\underline{s}$. 
The large $|\mu|$ behavior of $\Phi_q(i\mid\mu)$ 
can be computed as follows;  
\begin{eqnarray}
&&
\Phi_q(i\mid \mu)
\nonumber \\[1.5mm]
&&
=
\displaystyle{ 
 \sum_{j=i+1}^{+\infty}\,
 \ln \left(q^{\frac{1}{2}\left(i-j\right)}
           -q^{\frac{1}{2}\left(j-i\right)}\right)
 -\sum_{j \neq i}^{+\infty}\,
 \ln \left(
    q^{\frac{1}{2}\left(x_j(\mu)-x_i(\mu)\right)}
    -q^{\frac{1}{2}\left(x_i(\mu)-x_j(\mu)\right)}
     \right)}
\nonumber \\[1.5mm]
&&
=
\displaystyle{
  \frac{N}{\hbar}
  \int_{0}^{+\infty}dx \,
  \ln \Bigl(e^{Rx}-e^{-Rx}\Bigr)
  -\frac{N}{\hbar}
  \int_{0}^{+\infty}ds \, 
  \ln \Bigl(
      e^{R(u(\underline{s}\mid\,\mu)-u(s\mid\,\mu))}
      -e^{R(u(s\mid\,\mu)-u(\underline{s}\mid\,\mu))}
      \Bigr)}
\nonumber \\*
&&
~~+O(\hbar^0)
\nonumber \\[1.5mm]
&&
=
\displaystyle{
   \frac{N}{\hbar}
   \int_0^{+\infty}dx\, 
   \ln \Bigl( \frac{\sinh Rx}{R \Lambda}\Bigr)
   -\frac{N}{\hbar}
   \int_0^{+\infty}ds\, 
   \ln \left\{
         \frac{\sinh R(u(\underline{s}\mid\mu)-u(s\mid \mu))}
              {R \Lambda}
       \right\}}
\nonumber \\*
&&
~~+O(\hbar^0),  
\label{asymptotic Phi}
\end{eqnarray}
where the last expression enables us  
to make contact with the four-dimensional theory.  
By plugging (\ref{asymptotic K}) and (\ref{asymptotic Phi})
into (\ref{chi star 2}), 
the large $|\mu|$ behavior of $\chi_q^*(z\mid\mu)$ can 
be read as 
\begin{eqnarray}
&&
\chi_q^*(z\mid \mu)
\approx
C\, \frac{N}{\hbar}
\int_0^{+\infty} d\underline{s}
\,\,z^{-\frac{N}{\hbar}u(\underline{s}\mid\, \mu)}
\nonumber \\*[1.5mm]
&&
~~~~~~~
\times 
\exp \left\{
-\frac{RN}{2\hbar}u(\underline{s}\mid \mu)^2
+\frac{N}{\hbar}
\int_{-\infty}^{+\infty}
dv \,
\frac{ds(v\mid \mu)}{dv}
\ln 
\left(
\frac{\sinh R(u(\underline{s}\mid\mu)-v)}{R\Lambda}
\right)
\right\}, 
\nonumber \\*
\label{asymptotic chi star 2}
\end{eqnarray}
where 
$C \equiv 
\exp \left\{ 
\frac{N}{\hbar}
\int_0^{+\infty} dx 
\ln \left(\frac{\sinh Rx}{R\Lambda}\right)
\right\}$. 
The constant $C$ will be dropped out by absorbing it 
into $\psi^*(z)$ as a wave function renormalization. 
The density function appears in the above  
by the change of variable, $v=u(s\mid\mu)$. 
Thus we obtain the asymptotics (\ref{asymptotic chi star}).

The large $|\mu|$ behavior of $\chi_q(z\mid\mu)$ 
can be computed in the similar manner as above 
by noting the relation (\ref{chi vs chi star}). 
It becomes 
\begin{eqnarray}
&&
\chi_q(z\mid \mu)
\approx
C^{-1}
\frac{N}{\hbar}
\int_0^{+\infty}
d\underline{s}
~z^{-\frac{N}{\hbar}u(\underline{s}\mid\, \widetilde{\mu})}
\nonumber \\*[1.5mm]
&&
~~~~~~~
\times 
\exp \left\{
\frac{RN}{2\hbar}u(\underline{s}\mid \widetilde{\mu})^2
-\frac{N}{\hbar}
\int_{-\infty}^{+\infty}
dv\, 
\frac{ds(v\mid \mu)}{dv}
\ln 
\left(
\frac{\sinh R(-u(\underline{s}\mid\widetilde{\mu})-v)}{R\Lambda}
\right)
\right\}.  
\nonumber \\
\label{asymptotic chi 2}
\end{eqnarray}
The constant $C^{-1}$ will be dropped out. 
In the above, the density function of $\mu$ has been used 
instead of $\widetilde{\mu}$. These two are related by 
\begin{eqnarray}
\frac{ds}{du}(u\mid\widetilde{\mu})=
-1+\frac{ds}{du}(-u\mid\mu). 
\label{density reflection formula}
\end{eqnarray}

\section{Energy functions of partition}

The Boltzmann weight for a partition $\mu$ is 
$Q^{|\mu|}e^{V(\mu)}s_{\mu}(q^{-\rho})^2$. 
The Boltzmann weight for a very large partition 
can be measured by the so-called energy function.  
The technique used in the previous section 
is also relevant to obtain the energy function.

Let $\mu$ be a partition of order $\hbar^{-2}$. 
We rescale the Maya diagram $x_i(\mu)$ to 
$u(s\mid\mu)$ by (\ref{u(s)}).  
The asymptotics of the Boltzmann weight   
follows from the large $|\mu|$ behaviors of 
$s_{\mu}(q^{-\rho})$ and 
the $N$-periodic potential $V(\mu)$.  
The large $|\mu|$ behavior of 
$s_{\mu}(q^{-\rho})$ 
can be obtained by using (\ref{product formula a}).  
The computations analogous to the previous section 
give rise to  
\begin{eqnarray}
&&
\ln s_{\mu}(q^{-\rho})^2
\nonumber \\*[1.5mm]
&&
=\,
\frac{N^2}{\hbar^2}
\int_0^{+\infty}dx\, \,
x \ln \left(\frac{\sinh Rx}{R \Lambda}\right)^2 
-\frac{N^2}{\hbar^2}
\int_{-\infty}^{+\infty}du \,
\left( 1+2\frac{ds(u\mid\mu)}{du} \right)
\frac{Ru^2}{2}
\nonumber \\*[1.5mm]
&&
~~~
+\frac{N^2}{\hbar^2}
\int \int_{-\infty <u<v< +\infty} 
dudv \,
\left(1+\frac{ds(u\mid\mu)}{du}\right)
\frac{ds(v\mid\mu)}{dv}
\ln 
\left(
\frac{\sinh R(u-v)}{R \Lambda}
\right)^2 
\nonumber \\*[1.5mm]
&&
~~~
+O(\hbar^{-1}).
\label{asymptotic schur}
\end{eqnarray}

In order to make the asymptotics of 
the $N$-periodic potential be of order $\hbar^{-2}$ 
(the same order as (\ref{asymptotic schur})), 
we should regard $\xi_r$ in the potential 
as the order $\hbar^{-1}$ quantities and  
rescale them as well. 
Let $\zeta_r$ be the order $\hbar^0$ quantities 
defined by the rescalings 
\begin{eqnarray}
\xi_r=\frac{\zeta_r}{\hbar}, 
\label{zeta r}
\end{eqnarray}
where $\sum_{r=1}^N\zeta_r=0$. 
In Appendix A, we compute the asymptotics of 
the $N$-periodic potential 
in a way different from \cite{Nekrasov-Okounkov}. 
Our computations are based on the regularized density. 
To carry out these computations 
we need to make the following assumption; 
the density functions are non-decreasing for 
the partitions dominating near the thermodynamic limit.
As the result, we obtain 
\begin{eqnarray}
V(\mu)=
\frac{N}{\hbar^2}
\sum_{r=1}^{N}
\zeta_r
\int_{u_r}^{u_{r-1}}du\,\, u 
\frac{d^2s(u\mid\mu)}{du^2}
+O(\hbar^{-1}),
\label{asymptotic V}
\end{eqnarray}
where $u_r$ ($0 \leq r \leq N$) 
are defined by the conditions 
\begin{eqnarray}
\left.
\frac{d}{du}s(u\mid\mu)\,
\right|_{u=u_r}
=-\frac{r}{N}.
\label{ur}
\end{eqnarray}
Our assumption ensures that 
$u_r$ exist and are ordered 
$u_r < u_{r-1}$ with 
$u_{0,\,N}$$=\pm \infty$.

By using (\ref{asymptotic schur}) and (\ref{asymptotic V}) 
we can see that the Boltzmann weight asymptotes to
\footnote{
The statistical model considered in this paper 
is related \cite{quantum calabi-yau} 
to topological string on a non-compact toric 
Calabi-Yau threefold 
\cite{topological vertex}. 
The precise correspondence is described in 
\cite{MNTT}.  
$\hbar$ is identified with string coupling constant $g_{st}$.  
The asymptotic expansion corresponds to 
the perturbative string theory. 
The identification $q=e^{-\frac{2R\hbar}{N}}$
indicates that $\hbar$ appears in a pair with $1/N$ 
as observed in (\ref{asymptotic boltzmann}). 
This seems to indicate the correspondence with 
$1/N$ expansion of the underlying gauge theory.}
\begin{eqnarray}
Q^{|\mu|}e^{V(\mu)}s_{\mu}(q^{-\rho})^2
&\approx&
C'
Q^{|\mu|}
\exp \left\{
-\frac{N^2}{\hbar^2}
E \left[s(\cdot \mid\mu) \right]
\right\}, 
\label{asymptotic boltzmann} 
\end{eqnarray}
where $E\left[s(\cdot \mid\mu) \right]$ 
is the energy function of $\mu$ defined by
\begin{eqnarray}
&&
E\left[ s(\cdot \mid\mu) \right] 
\equiv 
\int_{-\infty}^{+\infty} du\, 
\left( 1+2\frac{ds(u\mid\mu)}{du} \right)
\frac{Ru^2}{2}
\nonumber \\*[1.5mm]
&&~~~~~
-\int \int_{-\infty <u<v< +\infty} 
dudv \,
\left(1+\frac{ds(u\mid\mu)}{du}\right)
\frac{ds(v\mid\mu)}{dv}
\ln 
\left(
\frac{\sinh R(u-v)}{R \Lambda}
\right)^2 
\nonumber \\*[1.5mm]
&&~~~~~
-\frac{1}{N}
\sum_{r=1}^{N}
\zeta_r
\int_{u_r}^{u_{r-1}}
du\, u 
\frac{d^2s(u\mid\mu)}{du^2},
\label{energy function} 
\end{eqnarray} 
and $C'\equiv 
\exp \left\{
\frac{N^2}{\hbar^2}
\int_0^{+\infty}
dx x \ln \left(\frac{\sinh Rx}{R \Lambda}\right)^2 
\right\}$. The constant $C'$ will be dropped out. 
At the four-dimensional limit the above energy function 
becomes 
\begin{eqnarray}
&&
E_{4d}\left[s(\cdot \mid\mu)\right]
\equiv 
\lim_{R \rightarrow 0}
E\left[ s(\cdot \mid\mu) \right] 
\nonumber \\*[1.5mm]
&&
~~~
=-\int \int_{-\infty <u<v< +\infty} 
dudv \, 
\left(1+\frac{ds(u\mid\mu)}{du}\right)
\frac{ds(v\mid\mu)}{dv}
\ln 
\left(
\frac{u-v}{\Lambda}
\right)^2
\nonumber \\*[1.5mm]
&&
~~~~~~
-\frac{1}{N}
\sum_{r=1}^{N}
\zeta_r
\int_{u_r}^{u_{r-1}}
du\, u 
\frac{d^2s(u\mid\mu)}{du^2}. 
\label{4d energy function}
\end{eqnarray}

The energy functions (\ref{energy function}) 
and (\ref{4d energy function}) are the functionals of $s(u\mid\mu)$. 
It is also possible to express them 
by using the so-called profile of a partition $\mu$.  
In Appendix B, we present the descriptions in terms of the profile 
and thereby compare  (\ref{energy function}) 
and (\ref{4d energy function}) with the energy functions 
used in \cite{Nekrasov-Okounkov}.  
The energy function (\ref{energy function}) turns out to be 
different from that used \cite{Nekrasov-Okounkov} 
for a description of five-dimensional supersymmetric 
$SU(N)$ Yang-Mills. 
The difference originates in the existence of 
the Chern-Simons term of five-dimensional gauge theories. 
It is shown \cite{MNTT} that   
the random plane partitions (\ref{Z SU(N)}) describe  
five-dimensional supersymmetric $SU(N)$ Yang-Mills 
with the Chern-Simons term at the thermodynamic limit. 
The Chern-Simons correction appears in (\ref{energy function}) 
as the $u^2$ potential term. 
This term vanishes at the four-dimensional limit. 
Therefore the energy function (\ref{4d energy function}) coincides 
with that used \cite{Nekrasov-Okounkov} for the 
description of four-dimensional $\mathcal{N}=2$ supersymmetric 
$SU(N)$ Yang-Mills.

\section{Semi-classical wave function and 
Seiberg-Witten differential}

In this section we investigate the thermodynamic limits 
of the one-point functions 
(\ref{Psi(lambda)star}) and (\ref{Psi(lambda)}). 
We use the asymptotics obtained in the previous sections. 
Near the thermodynamic limit 
(the $\hbar \rightarrow 0$ limit), 
partitions of order $\hbar^{-2}$ dominate 
in the statistical model.   
In order to describe the one-point functions at the 
thermodynamic limit, 
taking account of the expressions 
(\ref{Psi(lambda)star 2}) and (\ref{Psi(lambda) 2}), 
we need to find out the classical configurations 
from among these dominant partitions. 
As described in (\ref{asymptotic boltzmann}) 
their Boltzmann weights are measured 
by the energy function $E\left[s(\cdot\mid\mu)\right]$. 
This means that the classical configurations 
are the minimizers of $E\left[s(\cdot)\right]$.
Let us suppose $s_{\star}(u)$ be such a minimizer.

The thermodynamic limits or 
the semi-classical approximations  
of the fermion one-point functions prove to  
have a universal form in the sense that 
it is irrespective of detail of the minimizers. 
They turn out to be given by the following 
WKB type wave functions. 
\begin{eqnarray}
\Psi_q^*(z)&\approx& 
\exp \left\{
-\frac{N}{\hbar}
\int^{z}u(z)\,d \ln z 
\right\}, 
\label{Psi(lambda)star wkb}
\\[1.5mm]
\Psi_q(z)&\approx& 
\exp \left\{
\frac{N}{\hbar}
\int^{z}u(z)\,d \ln z 
\right\}.  
\label{Psi(lambda) wkb}
\end{eqnarray}
In the above, 
the function $u(z)$ is determined by the equation
\begin{eqnarray}
\frac{dS(u)}{du}=
N \ln z \,, 
\label{critical eq for 5d diagonal}
\end{eqnarray}
where 
\begin{eqnarray}
S(u)\equiv
-\frac{RN}{2}u^2
+N\int_{-\infty}^{+\infty}
dv\, \frac{ds_{\star}(v)}{dv}
\ln 
\left(
\frac{\sinh R(u-v)}{R\Lambda}
\right). 
\label{explicit classical action} 
\end{eqnarray}


~

Let us confirm (\ref{Psi(lambda)star wkb}) and 
(\ref{Psi(lambda) wkb}). 
By following the mode expansions of 
the complex fermions, 
we first write the fermion one-point functions as 
\begin{eqnarray}
\Psi_q^*(z)= 
\sum_{j=-\infty}^{+\infty}
\widehat{\Psi}_q^*(j)\,z^{-j},
~~~~~~
\Psi_q(j)= 
\sum_{j=-\infty}^{+\infty}
\widehat{\Psi}_q(j)\,z^{j-1}\,, 
\label{Psi(lambda) mode}
\end{eqnarray}
where 
$\widehat{\Psi}_q^*(j)$ and $\widehat{\Psi}_q(j)$ 
are respectively the one-point functions of 
$\psi^*_{j}$ and $\psi_{-j}$. 
In other words 
they are given by the contour integrals  
\begin{eqnarray}
\widehat{\Psi}_q^*(j)= 
\oint \frac{dz}{2\pi i}\,
z^{j-1} \, \Psi_q^*(z),
~~~~~~~
\widehat{\Psi}_q(j)=
\oint \frac{dz}{2\pi i}\,
z^{-j}\, \Psi_q(z)\, , 
\label{Psi(j)}
\end{eqnarray}
for each $j \in \mathbb{Z}$.

Tracing the representations 
(\ref{Psi(lambda)star 2}) and (\ref{Psi(lambda) 2})  
these wave functions can be expressed as the following 
statistical sums; 
\begin{eqnarray}
\widehat{\Psi}_q^*(j)&=&
\frac{1}{Z_q^{SU(N)}}
\sum_{\mu}
Q^{|\mu|}
e^{V(\mu)}
s_{\mu}(q^{-\rho})^2\,
\widehat{\psi}_q^*(j\mid \mu), 
\label{Psi(j)star 2}\\[1.5mm]
\widehat{\Psi}_q(j)&=&
\frac{1}{Z^{SU(N)}_q}
\sum_{\mu}
Q^{|\mu|}
e^{V(\mu)}
s_{\mu}(q^{-\rho})^2\,
\widehat{\psi}_q(j\mid \mu), 
\label{Psi(j) 2}
\end{eqnarray}
where $\widehat{\psi}^*_q(j\mid\mu)$ 
and $\widehat{\psi}_q(j\mid\mu)$ are obtained 
from $\chi_q^*(z\mid\mu)$ and 
$\chi_q(z\mid \mu)$ in 
(\ref{chi star}) and (\ref{chi}) by   
\begin{eqnarray}
\widehat{\psi}_q^*(j\mid\mu)
&=& 
\oint 
\frac{dz}{2\pi i}\,
z^{j-1}\,
\chi_q^*(z\mid\mu),
\label{hat psi star j}\\[1.5mm]
\widehat{\psi}_q(j\mid\mu)
&=& 
\oint 
\frac{dz}{2\pi i}\,
z^{-j}\,
\chi_q(z\mid\mu). 
\label{hat psi j}
\end{eqnarray}

The main contributions to $\chi_q^*(z\mid\mu)$ 
and $\chi_q(z\mid \mu)$  
come from $\widehat{\psi}_q^*(j\mid\mu)$ 
and $\widehat{\psi}_q(j\mid\mu)$ with $j \sim \hbar^{-1}$
when $\mu$ is a partition of order $\hbar^{-2}$. 
These are seen from 
(\ref{psi star mode on mu}) and (\ref{psi mode on mu}). 
It follows that $\psi_j^*$ generically 
subtracts $j$ boxes from the Young diagram 
and $\psi_{-j}$ adds $j$ boxes to it, 
when they act on the partition. 
In particular,  
$\widehat{\psi}_q^*(j\mid\mu)$ with $j \sim \hbar^{-1}$ 
subtracts the same number of boxes from the partition. 
It is also similar to 
$\widehat{\psi}_q(j\mid\mu)$.  
We rescale $j$ 
in (\ref{hat psi star j}) and (\ref{hat psi j}) 
to the order $\hbar^0$ quantity $u$ by  
\begin{eqnarray}
j=\frac{N}{\hbar}u. 
\label{j u}
\end{eqnarray}

Let us first consider the $\hbar$-expansions of 
$\widehat{\psi}_q^*(j\mid\mu)$ and $\widehat{\psi}_q(j\mid\mu)$, 
where we regard $j=O(\hbar^{-1})$ and $|\mu|=O(\hbar^{-2})$. 
The semi-classical terms of the $\hbar$-expansions 
can be obtained by plugging 
(\ref{asymptotic chi star}) and (\ref{asymptotic chi}) 
respectively into (\ref{hat psi star j}) and (\ref{hat psi j}) 
with putting $j=Nu/\hbar$. 
They turn out to be 
\begin{eqnarray}
\widehat{\psi}^*_q(j\mid\mu) 
&\approx& 
\exp\left\{
\frac{1}{\hbar}
S \left[\,u;s(\cdot\mid\mu)\right]\right\}, 
\label{asymptotic hat star psi(j:mu)} 
\\[1.5mm]
\widehat{\psi}_q(j\mid\mu) 
&\approx&
\exp\left\{
\frac{-1}{\hbar}
S \left[\,u;s(\cdot\mid\mu)\right]\right\},  
\label{asymptotic hat psi(j:mu)}
\end{eqnarray}
where 
\begin{eqnarray}
S \left[\,u;s(\cdot\mid\mu)\right]=
-\frac{RN}{2}u^2
+N\int_{-\infty}^{+\infty}
dv\, \frac{ds(v\mid\mu)}{dv}
\ln 
\left(
\frac{\sinh R(u-v)}{R\Lambda}
\right). 
\label{S[u;s]}
\end{eqnarray}

The semi-classical approximations of 
$\widehat{\Psi}_q^*(j)$ and $\widehat{\Psi}_q(j)$ 
can be obtained as follows. 
Let $s_{\star}(u)$ be a minimizer 
of the energy function $E\left[s(\cdot)\right]$.
The representations (\ref{Psi(j)star 2})
and (\ref{Psi(j) 2}) will be helpful. 
Taking account of these statistical representations, 
their semi-classical approximations are given by 
(\ref{asymptotic hat star psi(j:mu)}) and  
(\ref{asymptotic hat psi(j:mu)}) 
evaluated at the minimizer $s_{\star}(u)$.  
Thus we obtain 
\begin{eqnarray}
\widehat{\Psi}_q^*(j)
&\approx& 
\exp 
\left\{ \frac{1}{\hbar}S(u) \right\}, 
\label{Psi(u)star wkb}
\\[1.5mm]
\widehat{\Psi}_q(j)
&\approx& 
\exp 
\left\{\frac{-1}{\hbar}S(u) \right\}, 
\label{Psi(u) wkb}
\end{eqnarray}
where the classical action $S(u)$ is read from 
(\ref{asymptotic hat star psi(j:mu)}) and 
(\ref{asymptotic hat psi(j:mu)}) as 
\begin{eqnarray}
S(u)= 
S \left[\,u; s_{\star}(\cdot) \right] 
\label{classical action}
\end{eqnarray}
and given by (\ref{explicit classical action}).

The semi-classical approximations or 
the thermodynamic limits of $\Psi_q^*(z)$ and $\Psi_q(z)$ 
can be obtained from (\ref{Psi(u)star wkb})
and (\ref{Psi(u) wkb}). 
Taking account of the scaling (\ref{j u}) 
we can replace the expressions (\ref{Psi(lambda) mode})  
with the following integrals at the thermodynamic limit. 
\begin{eqnarray}
\Psi_q^*(z)&\approx&
\frac{N}{\hbar}
\int_{-\infty}^{+\infty}du \,
\exp \left\{
\frac{1}{\hbar}
\left(
-u \ln z^N+S(u) 
\right)
\right\}, 
\label{Psi(lambda) star int}
\\[1.5mm]
\Psi_q(z)&\approx& 
\frac{N}{\hbar}
\int_{-\infty}^{+\infty}du \,
\exp \left\{
\frac{1}{\hbar}
\left(
u \ln z^N-S(u) 
\right)
\right\}.  
\label{Psi(lambda) int}
\end{eqnarray}
Since $\hbar$ is very small, 
we can evaluate the above integrations  
by applying the saddle point method. 
Critical points of the exponents become  
solutions of the equation 
(\ref{critical eq for 5d diagonal}).
Let $u(z)$ be the critical point. 
The critical value can be obtained as follows 
\begin{eqnarray}
u(z)\ln z^N-S(u(z))
&=&
\int^{u(z)}
d \left(u\frac{dS(u)}{du}\right) 
-\int^{u(z)}
du \frac{dS(u)}{du}
\nonumber \\[1.5mm]
&=&
\int^{u(z)}
u d \left(\frac{dS(u)}{du}\right)
\nonumber \\[1.5mm]
&=&
N\int^{z}
u(z)\,d\ln z. 
\label{critical value for 5d diagonal}
\end{eqnarray}
Therefore, the saddle point method gives rise 
to the semi-classical wave functions 
of the one-point functions at the main diagonal 
as described in (\ref{Psi(lambda)star wkb}) and 
(\ref{Psi(lambda) wkb}).

\subsubsection*{Semi-classical wave function 
and Seiberg-Witten differential}

Roles of the semi-classical wave functions 
(\ref{Psi(lambda)star wkb}) and (\ref{Psi(lambda) wkb})
in the Seiberg-Witten geometry are not obvious. 
In order to get some insight we look 
at the four-dimensional case.  
Let us first supplement the previous discussions by 
providing the four-dimensional limits of the relevant 
quantities. The four-dimensional limits of 
(\ref{asymptotic hat star psi(j:mu)}) and 
(\ref{asymptotic hat psi(j:mu)}) become  
\begin{eqnarray}
\widehat{\psi}_{4d}^{*}(j\mid\mu) 
\,\equiv\, 
\lim_{R \rightarrow 0}
\widehat{\psi}_q^*(j\mid\mu)
&\approx& 
\exp\left\{
\frac{1}{\hbar}
S_{4d} \left[\,u;s(\cdot\mid\mu)\right]\right\},
\label{asymptotic 4d hat star psi(j:mu)}
\\[1.5mm]
\widehat{\psi}_{4d}(j\mid\mu) 
\,\equiv\, 
\lim_{R \rightarrow 0}
\widehat{\psi}_q^*(j\mid\mu)
&\approx& 
\exp\left\{
\frac{-1}{\hbar}
S_{4d} \left[\,u;s(\cdot\mid\mu)\right]\right\}, 
\label{asymptotic 4d hat psi(j:mu)}
\end{eqnarray}
where 
\begin{eqnarray}
S_{4d}\left[\,u;s(\cdot\mid\mu) \right]=
N\int_{-\infty}^{+\infty}
dv\, \frac{ds(v\mid\mu)}{dv}
\ln 
\left(
\frac{u-v}{\Lambda}
\right). 
\label{4d S[u;s]}
\end{eqnarray}
Thus the minimizer $s_{\star}(u)$ of 
$E_{4d}\left[s(\cdot)\right]$ leads to  
the following semi-classical wave functions.   
\begin{eqnarray}
\widehat{\Psi}_{4d}^*(j)\,\equiv\, 
\lim_{R \rightarrow 0}
\widehat{\Psi}_q^*(j)
&\approx& 
\exp 
\left\{ \frac{1}{\hbar}S_{4d}(u) \right\}, 
\label{4d Psi(u)star wkb}
\\[1.5mm]
\widehat{\Psi}_{4d}(j)\,\equiv\, 
\lim_{R\rightarrow 0}
\widehat{\Psi}_q(j)
&\approx& 
\exp 
\left\{\frac{-1}{\hbar}S_{4d}(u) \right\}, 
\label{4d Psi(u) wkb}
\end{eqnarray}
where 
\begin{eqnarray}
S_{4d}(u)=
S_{4d} \left[\,u; s_{\star}(\cdot) \right]. 
\label{4d classical action}
\end{eqnarray}

We use the solution obtained by Nekrasov and Okounkov
\cite{Nekrasov-Okounkov}. 
Let ${\cal C}_{N}$ be the spectral curve of the $N$-periodic 
Toda chain 
\begin{eqnarray}
h+\frac{1}{h}=\frac{P(x)}{\Lambda^N},  
\label{spectral curve N-toda}
\end{eqnarray}
where $P(x)$ is the $N$-th order monic polynomial. 
The curve ${\cal C}_N$ is identified 
\cite{Toda curve} with the Seiberg-Witten curve 
of ${\cal N}=2$ supersymmetric 
$SU(N)$ Yang-Mills \cite{ Seiberg-Witten,SU(N) SW curve}. 
The solution of \cite{Nekrasov-Okounkov} is expressed as 
\begin{eqnarray}
\int_{-\infty}^{+\infty}
dv \,
\frac{d^2s_{\star}(v)}{dv^2}
\frac{1}{x-v}=
\frac{1}{N}\frac{d \ln h}{dx}. 
\label{4d solution}
\end{eqnarray}
It follows from (\ref{4d solution}) that
the differential $dS_{4d}(x)$ equals to 
\begin{eqnarray}
dS_{4d}(x)
=dx \ln h. 
\label{dS}
\end{eqnarray}
Therefore, 
the classical action (\ref{4d classical action}) 
can be written as the following integral on ${\cal C}_N$; 
\begin{eqnarray}
S_{4d}(u)
=\int^u dx\,  \ln h. 
\label{4d classical action integral}
\end{eqnarray}

The thermodynamic limits of the four-dimensional 
counterparts of $\Psi_{q}^*(z)$ 
and $\Psi_{q}^*(z)$ are described by 
\begin{eqnarray}
\Psi_{4d}^*(z)
\,\equiv\,
\lim_{R \rightarrow 0}
\Psi_{q}^*(z)
&\approx&
\frac{N}{\hbar}
\int_{-\infty}^{+\infty}du \, 
\exp \left\{
\frac{1}{\hbar}
\left(
-u \ln z^N+S_{4d}(u) 
\right)
\right\}, 
\label{Psi(lambda) 4d star int}
\\[1.5mm]
\Psi_{4d}(z)
\,\equiv\, 
\lim_{R \rightarrow 0}
\Psi_{q}(z)
&\approx& 
\frac{N}{\hbar}
\int_{-\infty}^{+\infty}du \, 
\exp \left\{
\frac{1}{\hbar}
\left(
u \ln z^N-S_{4d}(u) 
\right)
\right\}.  
\label{Psi(lambda) 4d int}
\end{eqnarray}
We can also apply the saddle point method  
to the above integrations. 
In the present case 
equation for the critical point becomes 
\begin{eqnarray}
h=z^N. 
\label{critical eq for 4d diagonal}
\end{eqnarray}
We then obtain the following semi-classical 
wave functions.
\begin{eqnarray}
\Psi_{4d}^*(z)&\approx& 
\exp \left\{
-\frac{1}{\hbar}
\int^{z}x \,d \ln h 
\right\}, 
\label{Psi(lambda) 4d star wkb}
\\
\Psi_{4d}(z)&\approx& 
\exp \left\{
\frac{1}{\hbar}
\int^{z}x \,d \ln h 
\right\}.  
\label{Psi(lambda) 4d wkb}
\end{eqnarray}
Recall that $x d \ln h$ is nothing but 
the Seiberg-Witten differential $dS_{s.w}$
for the $SU(N)$ Yang-Mills. 
Therefore we have shown above 
that the semi-classical approximations 
or the thermodynamic limits of the fermion one-point 
functions are given by the WKB type wave functions 
whose classical action is the integral of $dS_{s.w}$ 
on the curve ${\cal C}_N$
\begin{eqnarray}
\Psi_{4d}^*(z)&\approx& 
\exp \left\{
-\frac{1}{\hbar}
\int^{z}dS_{s.w}
\right\}, 
\label{Psi(lambda) 4d star sw}
\\
\Psi_{4d}(z)&\approx& 
\exp \left\{
\frac{1}{\hbar}
\int^{z}dS_{s.w}
\right\}.  
\label{Psi(lambda) 4d sw}
\end{eqnarray}

Meromorphic function $h$ is the Floquet multiplier 
in the spectral analysis of the 
periodic Toda chain. 
Equation (\ref{critical eq for 4d diagonal}) shows that 
$z$ is nothing but the spectral parameter of the 
associated linear problem. 
This implies a hidden relation between 
the above fermion wave functions and the Baker-Akhiezer 
functions of the Toda hierarchy. 
It is argued in \cite{Takasaki-Nakatsu} that 
the isospectral problem (the $N$-band solitons) 
is converted to an isomonodoromy problem 
by imposing a homogeneity condition 
analogous to the renormalization group equation.  
The associated linear system turns to admit 
a WKB analysis (multiscale analysis of 
the isomonodoromy problem), 
which gives the semi-classical wave functions 
(\ref{Psi(lambda) 4d star sw}) and 
(\ref{Psi(lambda) 4d sw}).

\section{One-point functions away from the main diagonal}

$\Gamma(m)$ in the transfer matrix  
is the hamiltonian operator at the discretized time $m$. 
The discretized time becomes a continuous time $t$ 
at the $\hbar \rightarrow 0$ limit. 
In fact, 
$t$ is identified with a coordinate of the limit 
shape \cite{Okounkov-Reshetikhin} 
of random plane partition. 
The limit shape is interpreted 
\cite{quantum calabi-yau} 
as the mirror of the toric $\mathbb{C}^3$. 
Free fermions  
$\psi^*(z)$ and $\psi(z)$ 
have been proposed \cite{brane probe} as probe branes 
in the mirror. In this section we compute 
the semi-classical wave functions of the fermions 
located away from the main diagonal. 
We restrict to the $U(1)$ case. 
In particular, we put $q=e^{-2R\hbar}$. 
We compute these wave functions by applying 
the method developed in \cite{Okounkov-Reshetikhin}.  
It turns out that $t$ is identical to the scale parameter 
of the gauge theory.

One-point functions at the discretized time 
$m_0 \in \mathbb{Z}_{\leq 0}$ are given by 
\begin{eqnarray}
\Psi^*_q(z;\,m_0)
&\equiv& 
\frac{1}{Z^{U(1)}_q}
\langle \emptyset ;\, -1|
\Biggl\{
\prod_{m=-\infty}^{m_0-1}\Gamma(m)
\Biggr\}\,
\psi^*(z)\,
\Biggl\{
\prod_{m=m_0}^{-1}\Gamma(m)
\Biggr\}\,
Q^{L_0}\,
\Biggl\{
\prod_{m=0}^{+\infty}\Gamma(m)
\Biggr\}
|\emptyset ;\,0\rangle ,
\nonumber \\
&&
\label{Psi m0 star}
\\[1.5mm]
\Psi_q(z;\,m_0)
&\equiv&
\frac{1}{Z^{U(1)}_q}
\langle \emptyset;\, +1|
\Biggl\{
\prod_{m=-\infty}^{m_0-1}\Gamma(m)
\Biggr\}\,
\psi(z)\,
\Biggl\{
\prod_{m=m_0}^{-1}\Gamma(m)
\Biggr\}\,
Q^{L_0}\,
\Biggl\{
\prod_{m=0}^{+\infty}\Gamma(m)
\Biggr\}
|\emptyset;\,0\rangle .
\nonumber \\
&&
\label{Psi m0}
\end{eqnarray}
Note that one-point functions at a positive 
discretized time are defined in the same way as above,     
and that their computations become similar to those  
at the negative time presented below.

The above one-point functions can be calculated exactly. 
Let us define the adjoint actions of $\Gamma(m)$ 
on the fermions as  
\begin{eqnarray}
Ad\Bigl(\Gamma(m)\Bigr)\,
\psi^*(z)=
\Gamma(m)\,\psi^*(z)\,\Gamma(m)^{-1},~~~~
Ad\Bigl(\Gamma(m)\Bigr)\,
\psi(z)=
\Gamma(m)\,\psi(z)\,\Gamma(m)^{-1}. 
\label{def adjoint}
\end{eqnarray}
By using (\ref{def adjoint}) 
we can evaluate 
(\ref{Psi m0 star}) and (\ref{Psi m0}) as follows; 
\begin{eqnarray}  
\Psi^*_q(z;\,m_0)&=&
\langle \emptyset;\, -1|\,
Ad
\Bigl( 
\prod_{m=0}^{+\infty}\Gamma(m)^{-1}
\Bigr)
Ad
\left(
Q^{-L_0}
\right)
Ad
\Bigl( 
\prod_{m=-\infty}^{m_0-1}\Gamma(m)
\Bigr)
\psi^*(z)\,
|\emptyset;\,0 \rangle, 
\nonumber 
\\
\label{Psi m0 star 2}
\\[1.5mm]
\Psi_q(z;\,m_0)&=&
\langle \emptyset;\, +1|\,
Ad
\Bigl( 
\prod_{m=0}^{+\infty}\Gamma(m)^{-1}
\Bigr) 
Ad
\left( 
Q^{-L_0}
\right)
Ad
\Bigl( 
\prod_{m=-\infty}^{m_0-1}\Gamma(m)
\Bigr)
\psi(z)\,
|\emptyset;\,0 \rangle.   
\nonumber \\ 
\label{Psi m0 2}
\end{eqnarray}
The adjoint actions (\ref{def adjoint}) 
are computed from (\ref{gamma m}). 
They become 
\begin{eqnarray}
Ad \Bigl(\Gamma(m)\Bigr) 
\psi^*(z)
&=&
\left\{
\begin{array}{cl}
 \left(1-q^{m+\frac{1}{2}}z^{-1}\right) 
 \psi^*(z) 
&
~~~~~
\mbox{for}~~m \geq 0 \\[1.5mm]
 \left(1-q^{-m-\frac{1}{2}}z \right) 
 \psi^*(z) 
&
~~~~~
\mbox{for}~~m \leq -1,
\end{array}
\right.
\label{ad m psi star} \\[1.5mm]
Ad \Bigl(\Gamma(m)\Bigr) 
\psi(z)
&=&
\left\{
\begin{array}{cl}
  \left(1-q^{m+\frac{1}{2}}z^{-1} \right)^{-1}
  \psi(z) 
&
~~~~~
\mbox{for}~~m \geq 0 \\[1.5mm]
   \left(1-q^{-m-\frac{1}{2}}z \right)^{-1}
   \psi(z) 
&
~~~~~
\mbox{for}~~m \leq -1.
\end{array}
\right.
\label{ad m psi}
\end{eqnarray}
By plugging (\ref{ad m psi star}) and (\ref{ad m psi}) 
into (\ref{Psi m0 star 2}) and (\ref{Psi m0 2}) 
we obtain 
\begin{eqnarray}
\Psi^*_q(z;\,m_0)\,=\,
\frac{ 
   \Bigl(q^{-m_0+\frac{1}{2}}z;q \Bigr)_{\infty}}
{\Bigl(q^{\frac{1}{2}}Qz^{-1};q\Bigr)_{\infty}},
~~~~~~
\Psi_q(z;m_0)
\,=\,
\Psi^*_q(z;\,m_0)^{-1},  
\label{exact form}
\end{eqnarray}
where the infinite product 
$(a;q)_{\infty}\equiv \prod_{n=0}^{+\infty}(1-aq^n)$ 
is used.

\subsubsection*{Semi-classical wave functions}

Asymptotics of the higher modes of 
$\Psi_q^*(z;\,m_0)$ and $\Psi_q(z;\,m_0)$ 
can be studied as in accord with the previous discussion 
at the main diagonal. 
Let $j \in \mathbb{Z}$. 
We introduce 
\begin{eqnarray}
\widehat{\Psi}_q^*(j;\,m_0)
&\equiv&
\oint \frac{dz}{2\pi i}\,
z^{j-1}\, 
\Psi^*_q(z;m_0), ~~~~~
\label{psi star j m0} \\
\widehat{\Psi}_q(j;\,m_0)
&\equiv&
\oint \frac{dz}{2\pi i}\,
z^{-j}\,
\Psi_q(z;m_0)  . 
\label{psi j m0}
\end{eqnarray}

To find out their asymptotics 
we need to rescale the integers $j$ and $m_0$ appropriately. 
Guided by (\ref{j u}), 
we regard $m_0$ also of order $\hbar^{-1}$ 
and introduce the order $\hbar^0$ quantities $u$ and $t$ by 
\begin{eqnarray}
j=\frac{u}{\hbar},~~~~~~~~
m_0=\frac{t}{\hbar}, 
\label{u and t}
\end{eqnarray}
where $u \in \mathbb{R}$ and $t \in \mathbb{R}_{\leq 0}$. 
Let us consider the $\hbar$-expansions of 
$\widehat{\Psi}^*_q\left(\frac{u}{\hbar};\frac{t}{\hbar}\right)$
and 
$\widehat{\Psi}_q\left(\frac{u}{\hbar};\frac{t}{\hbar}\right)$ 
($q=e^{-2R\hbar}$). 
They turn out to be written in the following form.  
\begin{eqnarray}
\widehat{\Psi}_q^*
\left(\frac{u}{\hbar}\,;\,\frac{t}{\hbar}\right)
&=&
\oint \frac{dz}{2\pi i}\,
\exp \left\{ 
\frac{1}{\hbar}S^{(0)}(z;(u,t))
+
O(\hbar^0)
\right\}, 
\label{psi(u,t) star}
\\[1.5mm]
\widehat{\Psi}_q
\left(\frac{u}{\hbar}\,;\,\frac{t}{\hbar}\right)
&=&
\oint \frac{dz}{2\pi i}\,
\exp \left\{ 
-\frac{1}{\hbar}S^{(0)}(z;(u,t))
+
O(\hbar^0)
\right\}.  
\label{psi(u,t)}
\end{eqnarray}
The classical action $S^{(0)}(z;(u,t))$ 
in the above can be computed as follows. 
We first notice that the $\hbar$-expansion of 
$\ln (a;q)_{\infty}$ reads as 
\begin{eqnarray}
\ln (a;q)_{\infty}=
\frac{1}{2R\hbar}
\int_0^a dx \,
\frac{\ln (1-x)}{x}
+O(\hbar^0). 
\label{hbar expansion ln(a;q)}
\end{eqnarray}
Then, (\ref{exact form}) leads to  
\begin{eqnarray}
&&
\ln z^{j-1} \,\Psi_q^*(z;\,m_0) 
\nonumber \\*[1.5mm]
&&
~~~
=(j-1) \ln z 
+\ln 
\left(q^{-m_0+\frac{1}{2}}z;q \right)_{\infty}
-\ln 
\left(q^{\frac{1}{2}}Qz^{-1};q\right)_{\infty} 
\nonumber \\[1.5mm]
&&
~~~
=
\frac{1}{\hbar}
\left\{ 
u \ln z 
+\frac{1}{2R}\int_0^{z}dx \,  
\frac{\ln \left( 1-e^{2Rt}x \right)}{x}
-\frac{1}{2R}\int_0^{z^{-1}}dx \,  
\frac{\ln \left(1-Qx \right)}{x}
\right\} 
\nonumber \\
&&
~~~~~
+
O(\hbar^0). 
\label{asymptotic ln psi star}
\end{eqnarray}
Therefore we obtain 
\begin{eqnarray}
&&
S^{(0)}(z;(u,t))
\nonumber \\*[1.5mm]
&&
~~~
=
u\ln z 
+\frac{1}{2R}\int_0^{z}dx \,
\frac{\ln \left( 1-e^{2Rt}x \right)}{x}
-\frac{1}{2R}\int_0^{z^{-1}}dx \,  
\frac{\ln \left(1-Qx \right)}{x}. 
\label{S(lambda;(u,t))}
\end{eqnarray}

Since $\hbar$ is very small,  
the saddle point method becomes applicable  
to the contour integrals 
in (\ref{psi(u,t) star}) and (\ref{psi(u,t)}). 
The critical points of $S^{(0)}(z;(u,t))$ 
turn to be the solutions $\alpha, \bar{\alpha}$ 
of the quadratic equation 
\begin{eqnarray}
e^{2Rt}z^2
-(1+e^{2Rt}Q-e^{-2Rt})z
+Q=0. 
\label{quadratic eq}
\end{eqnarray}
This allows us to write the critical points as    
\begin{eqnarray}
\alpha,\bar{\alpha}=
Q^{\frac{1}{2}} e^{-Rt\pm i \theta_{\star}(u,t)}.  
\label{alpha}
\end{eqnarray}
The phase $\theta_{\star}(u,t)$ 
in (\ref{alpha}) satisfies the equation    
\begin{eqnarray}
\cos \theta_{\star}(u,t)=
R\Lambda_{eff}(t)+
\frac{1}{2\Lambda_{eff}(t)}
\frac{1-e^{-2Ru}}{2R}, 
\label{theta star equation}
\end{eqnarray}
where 
$\Lambda_{eff}(t)
\equiv  
\Lambda e^{Rt}$.

The saddle point method gives rise to 
the following semi-classical wave functions.  
\begin{eqnarray}
\widehat{\Psi}_q^*
\left(\frac{u}{\hbar}\,;\,\frac{t}{\hbar}\right)
&\approx& 
\exp \left\{
\frac{1}{\hbar}
S^{(0)}
(\alpha;(u,t)) \right\},~~~~~~
\label{psi star wkb} 
\\
\widehat{\Psi}_q 
\left(\frac{u}{\hbar}\,;\,\frac{t}{\hbar}\right)
&\approx&
\exp \left\{
\frac{-1}{\hbar}
S^{(0)}
(\alpha;(u,t)) \right\},
\label{psi wkb}
\end{eqnarray}
where the critical point $\alpha$ is taken 
for simplicity. The critical value
$S^{(0)}(\alpha;(u,t))$ can be read easily 
up to the $u$-independent term since we have 
\begin{eqnarray}
\frac{d}{du}S^{(0)}(\alpha;(u,t))=\ln \alpha.
\label{dS(0)/du}
\end{eqnarray}
The integration of (\ref{dS(0)/du}) gives rise to 
\begin{eqnarray}
S^{(0)}(\alpha;(u,t))
&=&
-Rtu
+\frac{u}{2}\ln Q  
+i\int^u dx\, \theta_{\star}(x,t)
\nonumber \\
&&
+ u\mbox{-independent term}.
\label{critical value}
\end{eqnarray}

\subsubsection*{Family of curves}

The above semi-classical analysis has its interpretation 
in complex geometry. To explain this, 
it is convenient to start with the four-dimensional case. 
It follows from (\ref{theta star equation}) that 
the phase $\theta_{\star}(u,t)$ becomes independent of 
$t$ at the limit $R \rightarrow 0$. 
Equation (\ref{theta star equation}) is translated to 
\begin{eqnarray}
\cos \theta_{\star}^{4d}(u)=\frac{u}{2\Lambda}, 
\label{4d theta star equation}
\end{eqnarray}
where  
$\theta_{\star}^{4d}(u) 
\equiv 
\lim_{R \rightarrow 0}
\theta_{\star}(u,t)$. 
The phase $\theta_{\star}^{4d}$ can be thought 
as a meromorphic function on a Riemann surface. 
Let $\mathcal{C}$ be the curve ($\mathbb{P}^1$) 
\begin{eqnarray}
y^2=x^2-4\Lambda^2,   
\label{4d U(1) curve}
\end{eqnarray}
and $h$ be the meromorphic function on 
$\mathcal{C}$ 
\begin{eqnarray}
h=\frac{x+y}{2\Lambda}. 
\label{4d U(1) h}
\end{eqnarray}
Then we can write equation (\ref{4d theta star equation}) as  
\begin{eqnarray} 
\theta_{\star}^{4d}(u)=
\left. \frac{1}{i}\ln h \, \right|_{x=u+i0}. 
\label{4d theta vs h}
\end{eqnarray}
The classical action becomes an integral 
of $dx \ln h$ on the curve $\mathcal{C}$. 
\begin{eqnarray}
i\int^udx \,\theta^{4d}_{\star}(x)=
\int^u dx \,\ln h. 
\label{4d action integral}
\end{eqnarray}
It is tempting to think $x$ and $\ln h$ in the above 
as the action-angle variables of the one-dimensional Toda chain.  
Note that the curve $\mathcal{C}$ can be expressed as 
\begin{eqnarray}
h+\frac{1}{h}=\frac{x}{\Lambda}. 
\label{spectral curve of infinite Toda}
\end{eqnarray} 
This is the spectral curve of the infinite Toda chain.

We now move on to the five-dimensional case. 
For each $t \in \mathbb{R}_{\leq 0}$, we associate 
the curve $\mathcal{C}_t$  
\begin{eqnarray}
y^2=z^2-4\Lambda_{eff}^2(t), 
\label{5d U(1) curve t}
\end{eqnarray}
and the meromorphic function $h_t$ 
\begin{eqnarray}
h_t\equiv \frac{y+z}{2\Lambda_{eff}(t)}. 
\label{5d U(1) ht}
\end{eqnarray} 
The identification of $u$ with $z$ gets 
slightly involved in the five-dimensional theory. 
Let $\mathbb{C}/\mathbb{Z}$ be 
the cylinder obtained by identifying $x$ 
with $x+ \frac{\pi i}{R}n$. 
We will regard $u$ as the real line of the cylinder. 
Note that the cylinder becomes $\mathbb{C}$ as $R \rightarrow 0$. 
We introduce the holomorphic function on $\mathbb{C}/\mathbb{Z}$ by 
\begin{eqnarray}
z_t(x)\equiv 
2R\Lambda_{eff}^2(t)+
\frac{1-e^{-2Rx}}{2R}. 
\label{z(x)t}
\end{eqnarray}
Then the equation (\ref{theta star equation}) 
can be written as 
\begin{eqnarray}
\theta_{\star}(u,t)=
\left.
\frac{1}{i}
\ln h_t \, 
\right|_{z=z_t(u+i0)}. 
\label{theta vs h}
\end{eqnarray}
This shows that $\theta_{\star}$ has the same form 
as the four-dimensional counterpart. 
Effect of the fifth dimension is encoded in $z_t(x)$ 
such that (\ref{theta vs h}) smoothly reduces to 
(\ref{4d theta vs h}) as $R \rightarrow 0$. 
In fact, we have 
$\lim_{R \rightarrow 0}z_t(x)=x$ and 
$\lim_{R \rightarrow 0}\Lambda_{eff}(t)=\Lambda$.  
Therefore (\ref{theta vs h}) becomes 
(\ref{4d theta vs h}) at the four-dimensional limit. 
The classical action becomes the following integral 
on the curve ${\cal C}_t$. 
\begin{eqnarray}
i\int^u dx \, \theta_{\star}(x,t) 
&=&
\int^u dx \, \ln h_t(z_t(x)) 
\nonumber \\
&=&
\int^{z_t(u+i0)}
dz 
\left(\frac{dz_t}{dx}\right)^{-1} \ln h_t. 
\label{classical action t}
\end{eqnarray}

An exact solution of the five-dimensional theory can be 
obtained from the geometrical data 
(\ref{5d U(1) curve t}), (\ref{5d U(1) ht}) 
and (\ref{z(x)t}) at $t=0$. 
It should be noted that the time dependence of 
these data is only via $\Lambda_{eff}(t)$. 
Recall that $\Lambda=\Lambda_{eff}(0)$ 
should be identified with the lambda parameter of the gauge theory. 
The standard dimensional argument shows that 
the renormalization group flow is realized effectively 
by scaling the lambda parameter  
although it is originally a RG invariant. 
We conjecturally identify the above time evolution 
of the system with the RG flow. 
Since $\Lambda_{eff}(t)=\Lambda e^{2Rt}$ ($t \leq 0$), 
it becomes zero at $t=-\infty$ and 
the curve (\ref{5d U(1) curve t}) 
gets degenerate to $y^2=z^2$. 
We expect that 
the geometrical data near $t=\pm \infty$, 
reflecting the holographic principle, 
describes the gauge theory in the perturbative regime  
\footnote{
The analysis of the one-point functions at a
positive time leads to the same geometrical data 
with a slight change of $\Lambda_{eff}(t)$. 
Namely, $\Lambda_{eff}(t)=\Lambda e^{-2Rt}$ 
for $t \geq 0$. 
We have $\lim_{t \rightarrow +\infty}\Lambda_{eff}(t)=0$. 
Therefore $\mathcal{C}_t$ becomes the same degenerate curve 
at $t=+\infty$. }. 
This issue will be reported elsewhere 
\cite{MNTT in progress} from the viewpoint of 
integrable systems.

\appendix 
\section{Proof of equation (\ref{asymptotic V})}

We first express 
the $N$-periodic potential (\ref{N-periodic potential}) 
in a form relevant to study its asymptotics. 
It is convenient to consider partitions 
paired with the $U(1)$ charges. 
Let $(\mu,n)$ be the charged partition, 
where $\mu$ is a partition and $n$ is the $U(1)$ charge. 
The states $|\mu ; n \rangle$ constitute bases of the Fock space 
of a single complex fermion.  
System of $N$ component fermions is realized 
\cite{Miwa-Jimbo} on this Fock space. 
Thereby any charged partition $(\mu,n)$ can 
be expressed in a unique way 
as a set of $N$ charged partitions $(\lambda^{(r)},p_r)$ 
and vice versa. In terms of the Maya diagrams 
the correspondence can be read as follows;  
\begin{eqnarray}
\Bigl\{ n+x_{i}(\mu) \Bigr\}_{i \geq 1}=
\bigcup_{r=1}^N~
\Bigl\{ N(p_r+x_{i_r}(\lambda^{(r)}))+r-1 
\Bigr\}_{i_r \geq 1}. 
\label{mu vs lambda(r)}
\end{eqnarray}
The periodic potential $V(\mu)$ is shown 
\cite{Nekrasov-Okounkov} to be $\sum_{r=1}^N\xi_rp_r$,  
where $p_r$ are the $U(1)$ charges of 
the $N$ charged partitions corresponding to $(\mu,0)$.

For a charged partition $(\mu,n)$, 
we introduce the density 
$\rho(x \mid \mu \,; n)$ by 
\begin{eqnarray}
\rho(x \mid \mu \,; n)\equiv
\sum_{i=1}^{+\infty}
\delta(x-n-x_{i}(\mu)). 
\label{rho(mu:n)}
\end{eqnarray}
In the neutral case we simply denote 
$\rho(x\mid\mu) \equiv \rho(x\mid \mu\,; 0)$. 
The density (\ref{rho(mu:n)}) is not sensitive to the 
$U(1)$ charge since $n$ could be absorbed into a shift 
of $x$. It is convenient to modify the above density as   
\begin{eqnarray}
\rho_{reg}(x\mid \mu\,; n)=
\sum_{i=1}^{+\infty}
\delta(x-n-x_{i}(\mu))
-\sum_{i=1}^{+\infty}\delta(x+i), 
\label{rho reg (mu,n)}
\end{eqnarray}
where the subtraction is prescribed so that it satisfies 
\begin{eqnarray}
\int_{-\infty}^{+\infty}dx\, \rho_{reg}(x\mid \mu\,; n)=n.
\label{reg condition}
\end{eqnarray}
The following expression of the $N$-periodic potential 
becomes important in the subsequent discussion. 
\begin{eqnarray}
V(\mu)=\sum_{r=1}^{N}\xi_r 
\int_{-\infty}^{+\infty}dx \,
\rho_{reg}(x\mid \lambda^{(r)}\,; p_r), 
\label{N-periodic potential 2}
\end{eqnarray}
where $(\lambda^{(r)},p_r)$ are the charged partitions 
which describe $(\mu,0)$. Note that $\sum_{r=1}^Np_r=0$.

When $\lambda^{(r)}$ are of order $\hbar^{-2}$ and  
$p_r$ are of order $\hbar^{-1}$, 
we rescale the variables as 
\begin{eqnarray}
x=\frac{u}{\hbar},~~~
i_r=\frac{s}{\hbar},~~~
p_r=\frac{\eta_r}{\hbar},~~~
x_{i_r}(\lambda^{(r)})
=\frac{u(s\mid\lambda^{(r)})}{\hbar}+O(\hbar^0).   
\label{rescale N pairs}
\end{eqnarray}
The asymptotics of the densities for $(\lambda^{(r)},p_r)$ 
as $\hbar \rightarrow 0$ can be computed 
in the similar manner as in the text. They become 
\begin{eqnarray}
\rho(x\mid \lambda^{(r)}\,; p_r)&=&
-\frac{ds(u-\eta_r\mid\lambda^{(r)})}{du}+O(\hbar^1), 
\label{rho vs ds}
\\
\rho_{reg}(x\mid \lambda^{(r)}\,; p_r)&=&
-\left\{
\frac{ds(u-\eta_r\mid\lambda^{(r)})}{du} 
+\theta(-u)
\right \}
+O(\hbar^1),  
\label{rho reg vs ds}
\end{eqnarray}
where $\theta(u)$ is the step function, that is, 
$\theta(u)=1$ for $u>0$, and $0$ for $u<0$.

Thanks to the correspondence (\ref{mu vs lambda(r)})  
we can write $\rho(x\mid\mu)$ and $\rho_{reg}(x\mid\mu)$ 
as the superpositions of 
the densities for the $N$ charged partitions 
in the following manner; 
\begin{eqnarray}
\rho(x\mid\mu)&=&
\frac{1}{N}\sum_{r=1}^N
\rho 
\Bigl(
\Bigl.
\frac{x-r+1}{N} \Bigr| \lambda^{(r)}\,; p_r 
\Bigr), 
\label{rho(mu) vs rho(lambda(r))} 
\\
\rho_{reg}(x\mid\mu)&=&
\frac{1}{N}\sum_{r=1}^N
\rho_{reg} 
\Bigl( \Bigl.
\frac{x-r+1}{N} \Bigr| \lambda^{(r)}\,; p_r
\Bigr). 
\label{rho(mu) vs rho(lambda(r)) reg} 
\end{eqnarray} 
(\ref{mu vs lambda(r)}) also shows that 
the rescalings (\ref{u and s}) and (\ref{u(s)}) 
are consistent with (\ref{rescale N pairs}). 
Therefore, at the thermodynamic limit, 
the above relations turn to be  
\begin{eqnarray}
\frac{ds(u\mid\mu)}{du}=
\frac{1}{N}\sum_{r=1}^N
\frac{ds(u-\eta_r\mid\lambda_r)}{du},  
\label{ds(mu) vs ds(lambda(r))}
\end{eqnarray}
where $\sum_{r=1}^N\eta_r=0$.

$\frac{d}{du}s(u\mid\mu)$ takes values in $[-1,0]$ 
and asymptotes respectively to $0$ 
as $u \rightarrow +\infty$ and $-1$ 
as $u \rightarrow -\infty$. 
At this stage we impose a condition on partitions. 
In the below our consideration is restricted to 
a class of partitions satisfying the condition 
that $\frac{d}{du}s(u\mid\mu)$ is non-decreasing. 
As shown in Appendix B, 
this is equivalent to say that the profile 
$f(u\mid\mu)$ is convex. 
For such a partition $\mu$, 
it follows from (\ref{reg condition}) 
and (\ref{rho reg vs ds}) that 
$\frac{d}{du}s(u-\eta\mid\mu)+\theta(-u)$ 
has a compact support.

Let us compute the asymptotics of $V(\mu)$ 
based on the expression (\ref{N-periodic potential 2}). 
We rescale $\xi_r$ in the potential 
to $\zeta_r$ by (\ref{zeta r}). 
The asymptotics can be computed 
by using (\ref{rho reg vs ds}) as 
\begin{eqnarray}
&&
\sum_{r=1}^{N}\xi_r 
\int_{-\infty}^{+\infty}dx \,
\rho_{reg}(x\mid \lambda^{(r)}\,; p_r) 
\nonumber \\[1.5mm]
&&
=\,
\frac{-1}{\hbar^2}
\sum_{r=1}^N
\zeta_r
\int_{-\infty}^{+\infty}
du \left\{
\frac{ds(u-\eta_r\mid\lambda^{(r)})}{du}
+\theta(-u) \right\}
+O(\hbar^{-1})
\nonumber \\[1.5mm]
&&
=\,
\frac{1}{\hbar^2}
\sum_{r=1}^N
\zeta_r
\int_{-\infty}^{+\infty}
du\,  u
\frac{d^2s(u-\eta_r\mid\lambda^{(r)})}{du^2}
+O(\hbar^{-1}), 
\label{N-periodic potential 3}
\end{eqnarray}
where the last equality follows 
by the partial integration.

\begin{figure}[ht]
\begin{center}
\includegraphics[scale=0.63]{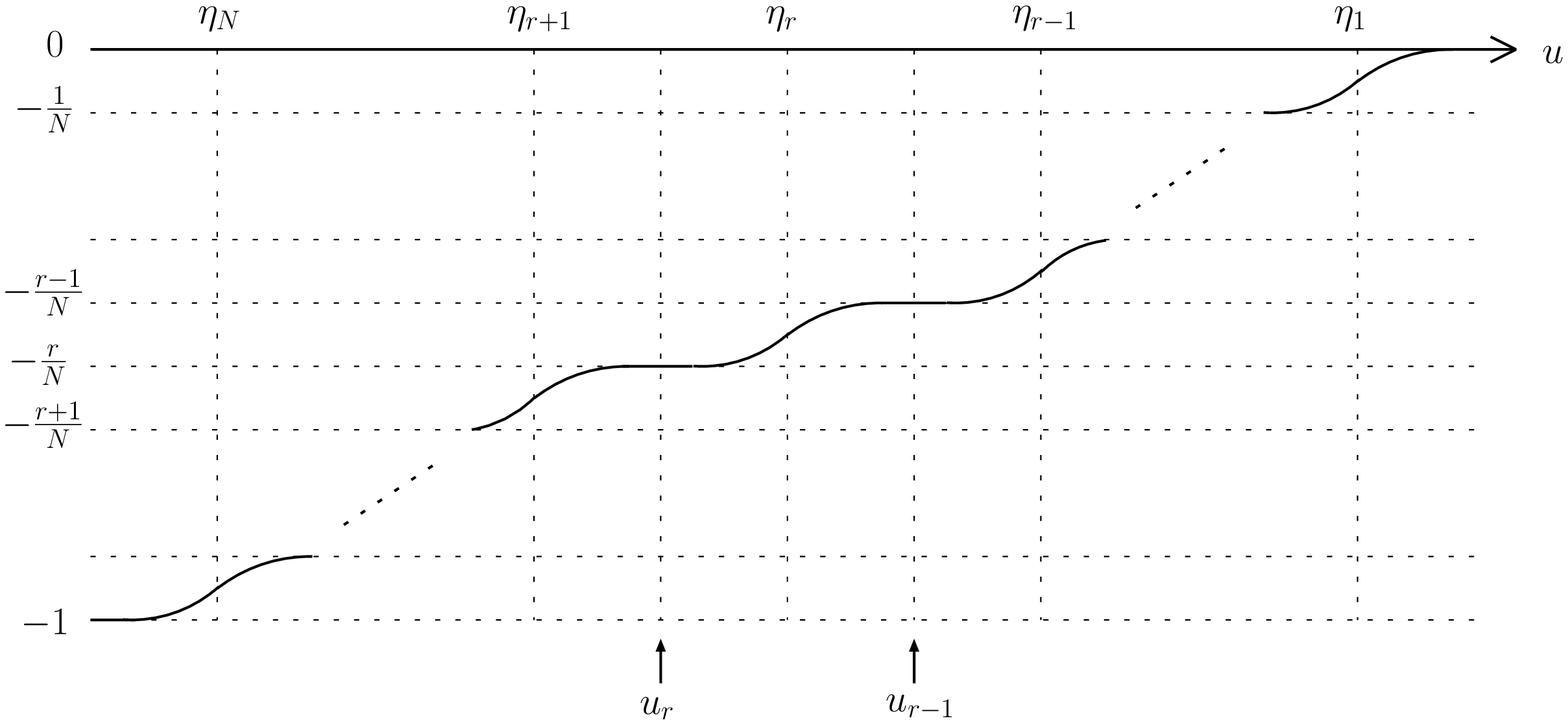}
\mycaption{3}{The graph of $\frac{d}{du}s(u\mid\mu)$ 
for $\eta_1>\eta_2>\cdots>\eta_N$, 
where $\eta_r$ are sufficiently separated from one another. 
$u_r$ ($0\leq r\leq N$) are determined by (\ref{ur}).}
\end{center}\label{fig2}
\end{figure}
Without losing generality it is enough to consider 
the case of $\eta_1>\eta_2> \cdots > \eta_N$. 
In addition, we suppose that $\eta_r$ are 
sufficiently separated from one another. 
The graph of $\frac{d}{du}s(u\mid\mu)$ is depicted in Figure 3. 
The relation (\ref{ds(mu) vs ds(lambda(r))}) leads to 
the following equalities; 
\begin{eqnarray}
\int_{-\infty}^{+\infty}
du\, u 
\frac{d^2s(u-\eta_r\mid\lambda^{(r)})}{du^2}=
N\int_{u_r}^{u_{r-1}}
du\,  u 
\frac{d^2s(u\mid\mu)}{du^2}, 
\label{equality}
\end{eqnarray}
where $u_r$ ($0 \leq r \leq N$) are determined 
by the condition (\ref{ur}).  
By plugging (\ref{equality}) into 
(\ref{N-periodic potential 3}) we obtain 
(\ref{asymptotic V}).

\section{Comparison between the energy functions of partitions}

We express the energy functions 
(\ref{energy function}) and (\ref{4d energy function}) 
by using the (rescaled) profile of partition. 
Let $\mu$ be a partition. The profile function 
$f_{\mu}(x)$ $(x \in \mathbb{R})$ 
is defined \cite{Nekrasov-Okounkov} by 
\begin{eqnarray}
f_{\mu}(x)
\equiv 
|x|+
\sum_{i=1}^{+\infty}
\Bigl \{
|x-x_i(\mu)-1|-|x-x_i(\mu)|
-|x+i-1|+|x+i|
\Bigr \}. 
\label{profile mu}
\end{eqnarray}
The profile function becomes a quantity of order $\hbar^{-1}$ 
when $\mu$ is a partition of order $\hbar^{-2}$.  
By the rescalings (\ref{u and s}) and (\ref{u(s)}) 
it is translated to 
\begin{eqnarray}
f_{\mu}(x)=
\frac{1}{\hbar}
f(u\mid\mu)+O(\hbar^0), 
\label{rescaled profile mu} 
\end{eqnarray} 
where the rescaled function $f(u\mid\mu)$ is described by 
\begin{eqnarray}
f(u\mid\mu)=
N|u|-2N
\int_0^{+\infty}
ds \Bigl \{ 
\theta(u-u(s\mid\mu))-\theta(u+s) 
\Bigr \}. 
\label{f vs u}
\end{eqnarray}
(\ref{f vs u}) leads to the following relation 
with the density function. 
\begin{eqnarray}
\frac{df(u\mid\mu)}{du}=
N 
\left( 
1+
2\frac{ds(u\mid\mu)}{du}
\right). 
\label{df vs ds}
\end{eqnarray}
We can rephrase the assumption made in Appendix A 
such that the rescaled profile functions for partitions 
dominating near the thermodynamic limit are convex.

The asymptotics (\ref{asymptotic V}) turns out to be 
the surface tension \cite{Nekrasov-Okounkov}. 
Let $\sigma(y)$ be the concave and piecewise-linear function 
on $[-N,N]$ defined by $\frac{d}{dy}\sigma(y)=\zeta_r$ for 
$y \in [N-2r,N-2(r-1)]$. 
It is a straightforward computation to see 
\begin{eqnarray}
N
\sum_{r=1}^{N}
\zeta_r
\int_{u_r}^{u_{r-1}}
du\, u 
\frac{d^2s(u\mid\mu)}{du^2}=
-\frac{1}{2}
\int_{-\infty}^{+\infty}du \,
\sigma \left( 
\frac{df(u\mid\mu)}{du} \right).
\label{surface tension}
\end{eqnarray}

The energy function 
(\ref{energy function}) 
can be rewritten as a functional of 
$f(u\mid\mu)$ by using (\ref{df vs ds}). 
It becomes  
\begin{eqnarray}
&& 
N^2
E \left[ s(\cdot \mid\mu) \right]
\,=\,
\int_{-\infty}^{+\infty} du\, 
\frac{df(u\mid\mu)}{du}
\frac{NRu^2}{2}
\nonumber \\*[1.5mm]
&&~~~~~
+
\frac{1}{4}
\int \int_{-\infty <u<v< +\infty} 
dudv \,
\left(N+\frac{df(u\mid\mu)}{du}\right)
\left(N-\frac{df(v\mid\mu)}{dv}\right) 
\ln 
\left(
\frac{\sinh R(u-v)}{R \Lambda}
\right)^2 
\nonumber \\*[1.5mm]
&&~~~~~
+\frac{1}{2}
\int_{-\infty}^{+\infty}du \,
\sigma \left( 
\frac{df(u\mid\mu)}{du} \right).
\label{energy function f}
\end{eqnarray}
Note that the above energy function is different from 
that used in \cite{Nekrasov-Okounkov} since  
the $u^2$ potential term does not appear there. 
While this, the four-dimensional limit 
(\ref{4d energy function}) becomes  
\begin{eqnarray}
&& 
N^2
E_{4d} \left[ s(\cdot\mid\mu) \right]
\nonumber \\*[1,5mm]
&&
~~~
=
\frac{1}{4}
\int \int_{-\infty <u<v< +\infty} 
dudv \,
\left(N+\frac{df(u\mid\mu)}{du}\right)
\left(N-\frac{df(v\mid\mu)}{dv}\right) 
\ln 
\left(
\frac{u-v}{\Lambda}
\right)^2
\nonumber \\*[1.5mm]
&&
~~~~~~
+\frac{1}{2}
\int_{-\infty}^{+\infty}du \,
\sigma \left( 
\frac{df(u\mid\mu)}{du} \right).
\label{4d energy function f}
\end{eqnarray}
This coincides with the energy function used  
in \cite{Nekrasov-Okounkov}.  
It is shown there that the minimizer of 
(\ref{4d energy function f}) 
is described by the Seiberg-Witten geometry 
of four-dimensional ${\cal N}=2$ supersymmetric $SU(N)$ 
Yang-Mills.

\subsection*{Acknowledgements}
T.N. is supported in part by Grant-in-Aid for 
Scientific Research 15540273. 
K.T. is supported in part by Grant-in-Aid for 
Scientific Research 16340040.


\end{document}